\documentclass[pra,aps,reprint,amsmath,amssymb,notitlepage,nofootinbib]{revtex4-1}
\usepackage[utf8]{inputenc}
\usepackage{hyperref,braket,graphicx,color,physics,mathtools}
\DeclarePairedDelimiter{\ceil}{\lceil}{\rceil}

\begin{document}
	
\title{Quantum information transfer using weak measurements and any non-product resource state} 
\author{Varad R.~Pande$^{1,2,3,4}$}
\email{varadrpande@gmail.com}
\author{Som Kanjilal$^2$}
\affiliation{$^1$Department of Physics, Indian Institute of Science Education and Research (IISER) Pune, India 411008,
\\
$^2$Center for Astroparticle Physics and Space Sciences, Bose Institute, Kolkata, India 695016,
\\
$^3$Harish-Chandra Research Institute, HBNI, Chhatnag Road, Jhunsi, Prayagraj (Allahabad), India 211019,
\\
$^4$Skolkovo Institute of Science and Technology (Skoltech), Bolshoy Boulevard 30, bld. 1 Moscow, Russia 121205.
}

\begin{abstract}
		
Information about an unknown quantum state can be encoded in weak values of projectors belonging to a complete eigenbasis. 
A protocol that enables one party -- Bob -- to remotely determine the weak values corresponding to weak measurements performed by another spatially separated party -- Alice is presented. The particular set of weak values contains complete information of the quantum state encoded on Alice's register, which enacts the role of preselected system state in the aforementioned weak measurement. Consequently, Bob can determine the quantum state from these weak values, which can also be termed as remote state determination or remote state tomography. A combination of non-product bipartite resource state shared between the two parties and classical communication between them is necessary to bring this statistical scheme to fruition. Significantly, the information transfer of a pure quantum state of any known dimensions can be effected even with resource states of low dimensionality and purity with a single measurement setting at Bob's end.

		
\end{abstract}
	
\maketitle

\section{Motivation}

\subsection{Quantum communication}
Quantum communication protocols such as randomness and measurement-based (BB84~\cite{bennett2014quantum}) or entanglement-based~\cite{einstein1935can,schrodinger1935discussion} (E91~\cite{ekert1991quantum}) quantum key distribution (M-QKD or E-QKD respectively), quantum teleportation~\cite{bennett1993teleporting} (QT), and remote state preparation~\cite{pati2000minimum,bennett2001remote} (RSP) allow secure data transmission beyond what is possible classically based on fundamental principles of quantum mechanics. Specifically, we rely on perfect randomness in choosing bits and measurement bases~\cite{bennett2014quantum}, uncertainty in the outcomes of (for the basic version) dichotomic projective quantum measurements~\cite{bennett2014quantum}, and quantum entanglement. The QKD protocols aim to distribute a secret key, which is a \emph{binary bit string} of certain length, between remote agents Alice and Bob who seek to communicate without an eavesdropper Eve getting information about it~\cite{scarani2009the}. In the M-QKD scheme, single photons are transmitted directly from Alice to Bob followed by a public certification process which establishes and verifies a secret key whose length is typically much lesser than the number of photons transmitted to begin with -- this ratio is termed as the secret key rate. In the E-QKD scheme, entangled photons are used by Alice and Bob to detect an eavesdropper and develop a secret key based again on a public certification process. QT and RSP seek to transmit \emph{quantum states} from Alice to Bob using a quantum entangled bipartite resource state and a classical communication channel, where the latter protocol requires lesser classical communication because knowledge of the state to be transmitted is known beforehand. The overarching goals of these methods are to establish large-scale quantum networks for secure communication, and to enable distributed quantum computation.

Several constraints must be overcome to achieve these goals. In teleportation (and RSP), we are restricted to qudit states (d-level quantum systems) and pure bipartite qudit resources, which has since been extended to teleport continuous variable quantum states~\cite{vaidman1994teleportation,braunstein1998teleportation,koniorczyk2001wigner,pirandola2015advances,yonezawa2004demonstration,andersen2013high} as well as to teleport from continuous to discrete variable quantum registers~\cite{PhysRevLett.118.160501}. These extensions and experimental implementations are conditional upon the dimensionality, purity, and amount of entanglement in the shared resource state~\cite{joo2003quantum,luo2013faithful,zhao2011faithful,albeverio2002optimal,PhysRevLett.119.110501,agrawal2002probabilistic}. A teleported quantum state does not provide classically useful information to conventional computers (the ``output problem"~\cite{biamonte2017quantum}) and tomography on a large number of copies teleported with high fidelity would be required if this is to succeed. For M-QKD, photon losses in the quantum channel restrict the communication distance and low efficiency of single-photon detectors is a roadblock in increasing the communication rate. For continuous variable QKD, the bandwidth of homodyne and heterodyne detectors and low electronic noise is a major issue. Generally, the heterodyne detectors are more sensitive to losses and detectors and optical fibres work in tandem for navigating the loss regime. The colder the detector is, the more loss it can tolerate in the fibre~\cite{diamanti2016practical}. Another source of error in the E-QKD scheme is the impure entanglement shared between distant parties. Here the detector efficiency is also affected by the coupling efficiencies of a parametric down conversion process subject to crystal nonlinearity. Additionally, background events of the coincidence counts reduce the probability of a genuine concidence event corresponding to a pump pulse~\cite{ma2007quantum}, which affects the communication rate. To increase E-QKD distance, quantum repeaters are needed for entanglement swapping and distillation among multiple distinct photon pairs. As such, many trusted nodes are required for this to succeed over intercontinental distances~\cite{lo2014secure}. Another peculiarity of E-QKD is its low key-generation rate in the low and medium loss regimes~\cite{ma2007quantum,lo2014secure}.

\subsection{Direct tomography using weak values}
Weak values proposed in a seminal paper~\cite{aharonov1988result} are complex entities which appear as a shift in the expectation value of a pointer observable when a weak von Neumann interaction between the system (observable $ \hat{A} $) and pointer states is followed by post selection on the system state~\cite{PhysRevD.40.2112,kanjilal2016manifestation}:
\begin{equation}
		\langle \hat{A} \rangle_w = \frac{\bra{\psi_f}A{\ket{\psi_i}}}{\braket{\psi_f}{\psi_i}},
\end{equation}
where $ \ket{\psi_f} $ and $ \ket{\psi_i} $ are the final and initial system states respectively.
Although weak measurements were initially introduced in the context of continuous variable Gaussian pointer states, the paradigm has since been theoretically and experimentally established for qubit pointer states~\cite{wu2009weak,brun2008test,lundeen2005practical}, entangled pointer states~\cite{menzies2008weak,ho2016experimental} and most generally to arbitrary pointer states~\cite{johansen2004weak,kofman2012nonperturbative}.
The concept of weak values and measurement has led to the development of ingenious methods for direct determination of a quantum wave-function and density matrix of mixed states followed by their experimental demonstration~\cite{lundeen2011direct,lundeen2012procedure,PhysRevLett.117.120401,wu2013state}. These schemes enable the state of a quantum system being probed appear directly as a shift in the expectation value of the pointer observable in terms of the weak values without the complicated state reconstruction process in conventional tomography~\cite{vogel1989determination,cramer2010efficient}. For a continuous variable pure state expressed as a quantum wave-function $ \psi(x) $, the wave-function at a particular position is equal to the weak value of position observable $ \ket{x_0}\bra{x_0} $ obtained after post-selection on a zero momentum eigenstate~\cite{lundeen2011direct}. Measuring a number of these weak values for several position eigenkets is thus enough to faithfully approximate the wave-function. A discrete variable pure state defined on a $d$-dimensional Hilbert space can be expressed in terms of vectors of a $d$-element orthonormal basis set: $ \ket{\psi}=\Sigma_{k=0}^{d-1} \braket{a_{k}}{\psi}\ket{a_{k}} $. Given such a set, it is possible to find another basis set which is mutually unbiased with respect to the former and has an element $ \ket{b_0} $ which is a discrete analogue ($ \ket{+} $ state corresponding to the d-dimensional Hadamard transform of $ \ket{0} $) of the zero momentum eigenstate in the continuous basis. This element satisfies $\braket{b_{0}}{a_{k}}=1 /\sqrt{d}$ $ \forall $ $k$. Thus, one can write:
\begin{equation}
	\label{stwv}
	\braket{a_{k}}{\psi}=\sqrt{d}\braket{b_{0}}{\psi}\frac{\braket{b_{0}}{a_{k}}\braket{a_{k}}{\psi}}{\braket{b_{0}}{\psi}}=\sqrt{d}\braket{b_{0}}{\psi} ( \hat{\Pi}^{k} )_{w},
\end{equation}
where $\hat{\Pi}^{k}$  is the projector corresponding to $\ket{a_{k}}$. The state can be rewritten in terms of these weak values: $ \ket{\psi}=\sqrt{d}\braket{b_{0}}{\psi}\Sigma_{k=0}^{d-1} ( \hat{\Pi}^{k} )_{w}\ket{a_{k}} $. 

\subsection{Objective}

We shall seek solutions to the quantum communication challenges by effectively combining broad characteristics of the above two seemingly distinct quantum information theoretic schemes and devise a non-local scenario of weak measurement to accomplish remote state determination (RSD) or remote state tomography (RST). RSD can transfer the information of an unknown pure quantum state $ \ket{\psi}_I $ of any known dimensions, by encoding its amplitudes in a complete set of weak values, from one party, Alice, to another spatially separated party, Bob, with any shared non-product resource state $ \rho_{AB} $ using local operations and classical communication. This attempts a simultaneous resolution to the threefold issues of resource dimensionality, purity, and amount of entanglement in protocols such as E-QKD, teleportation, and RSP which require pure entanglement. Therefore, it could principally resolve the issue of distance affecting these methods. Further, it can be compared with M-QKD, which requires the single-photons to be in an almost pure state when they reach Bob, detected by him with high probability and whose distance is affected by photon losses and noise in the quantum channel. Thus, it could solve the following problems: (i) it could be used as a method to establish a secret key between very distant observers while maintaining a reasonable data transmission rate by encoding the key in amplitudes of the state whose information is transferred; (ii) it could be used to transmit high-dimensional quantum states for distributed quantum information processing on qudits which otherwise requires high-dimensional pure entanglement shared over long distance to achieve using teleportation; and (iii) it could be used as a secure method to directly transmit any information (without establishing a secret key) encoded in amplitudes of the quantum state being transmitted. 
It (along with its possible variants) could also be used as a single solution to more than one of the three tasks within a broader quantum communication system.

We begin with a mathematical description of the protocol in Sec.~\ref{imaginary} and \ref{real} which enables the remote determination of a single weak value. Then we delineate the physical entities that need to be pre-decided in Sec.~\ref{pde}, the classical communication requirements in Sec.~\ref{cit} and the necessary and sufficient conditions to facilitate complete information transfer of any pure state in Sec.~\ref{nsc}. This is followed by a representative example in Sec.~\ref{example}, remarks on noise and error analysis in Sec.~\ref{noise}, and an outlook of this work in Sec.~\ref{conclusion}.

\section{Protocol}
\label{protocol}	
Alice and Bob share a bipartite non-product pure or mixed quantum state, $ \rho_{AB} $, which enacts the role of the resource. Alice has an system register on which the state $ \rho_{I} $, unknown to her, is encoded. As explained before, the state's probability amplitudes can be expressed as a set of weak values $ ( \hat{\Pi}^k_I )_w $ where index $k$ corresponds to one of the projectors belonging to its complete basis. In a single round of experimental runs, we seek to transfer one of these weak values. From hereon, we shall drop the index $k$. Alice begins by performing a weak interaction between her part ($ A $) of the shared state and system ($ I $) by letting them jointly evolve under the unitary $ \hat{U} = e^{ig\hat{\Pi}_I \otimes \hat{A} } $~\cite{von1955mathematical}. The total state after the weak interaction, characterized by expanding the coupling unitary up to the first order is $ \rho_{tw} \approx [\mathbb{I} + ig\hat{\Pi}_I \otimes \hat{A} \otimes \mathbb{I}] \rho_{I} \otimes \rho_{AB} [\mathbb{I} - ig\hat{\Pi}_I \otimes \hat{A} \otimes \mathbb{I}], $ where $\rho_{tw}$ is the total ($t$) post weak-interaction ($w$) state.
	
\subsection{Transferring imaginary part of the weak value}	
\label{imaginary}
We first derive the procedure with which Bob obtains imaginary part of the concerned weak value. To this end, Alice must perform a projective post-selection on $ \rho_I $ using the projector $ \ket{b_0}\bra{b_0} \equiv \hat{\pi}_{I}^v $. Index $ v $ denotes selection of the $v^{th}$ eigenvector of the chosen projection basis. After the post-selection, Bob's state can be obtained by tracing over parts $ A $ and $ I $ of the total state [see Der.~1 in Supplementary Materials]. Hence, the unnormalized state on Bob's side is $\rho_{B}^{un} = \Tr_{I,A} ( \hat{\pi}_{I}^v \rho_{tw})$: 
\begin{align}
\rho_B^{un} & \approx \Tr (\hat{\pi}_{I}^v \rho_{I}) ( ( \Tr_{A} ( \rho_{AB})  - ig  ( \hat{\Pi}_I )_w^* \Tr_{A} ( \rho_{AB} ( \hat{A} \otimes \mathbb{I})) \nonumber\\
& + ig ( \hat{\Pi}_I )_w \Tr_{A} ( (\hat{A} \otimes \mathbb{I}) \rho_{AB} ) )).
\end{align}
Here, we have used the definition of the complex weak value corresponding to the weak measurement performed by Alice between her part $ A $ of the shared state and the system $ I $ on which the state whose information is to be transferred is encoded: $ 	( \hat{\Pi}_I )_w \equiv \Tr ( \hat{\pi}_{I}^v \hat{\Pi}_I \rho_{I} ) / \Tr( \hat{\pi}_{I}^v \rho_{I} ).  $ It can be decomposed into its real and imaginary components: $ ( \hat{\Pi}_I )_w = \Re ( \hat{\Pi}_I )_w + i \Im ( \hat{\Pi}_I )_w $. When Bob measures the expectation value of an observable $ \hat{B} $ with respect to the normalized version of the above state [see Der.~\ref{A3} and \ref{A4} in \ref{appA}], it allows us to write the imaginary part of the weak value as
\begin{eqnarray}
\label{im}
\Im ( \hat{\Pi}_I )_w = \frac{\langle \hat{B} \rangle_f^{Im} - \Tr ( \hat{B} \rho_B^{in} ) }{\left( \splitfrac{2 g \bigg(\Tr ( (\hat{A} \otimes \mathbb{I}) \rho_{AB} ) \Tr ( \hat{B} \rho_B^{in} ) }{- \Tr ( \hat{B}  \Tr_{A}((\hat{A} \otimes \hat{B})\rho_{AB}) ) \bigg)} \right) }.
\end{eqnarray}
$ \langle \hat{B} \rangle_f^{Im} $ denotes the expectation value obtained by Bob on measuring his observable in this (first) set of experimental runs. $ \rho_{B}^{in} $ denotes Bob's initial reduced state. At this point, the expectation value with Bob has no information about the real part of $ ( \hat{\Pi}_I )_w $. Once he obtains the imaginary part of the weak value from the first set of experimental runs, Alice and Bob will proceed to the scheme for obtaining the real part.
	
\subsection{Transferring real part of the weak value}
\label{real}
In the $ 2^{nd} $ set of experimental runs, Alice changes the post-selection process. In addition to post-selecting on $ \rho_{I} $, she also post-selects on part $A$ of the shared state $ \rho_{AB} $ using the projector $ \hat{\pi}_{A}^l $ (index $ l $ denotes the $l^{th}$ eigenvector of the chosen projection basis) which does not commute with $ \hat{A} $. We can therefore write Bob's unnormalized final state [see Der.~\ref{B1} in \ref{appB}] by tracing over parts $ I $ and $ A $ corresponding to Alice's quantum registers: $ \rho_B^{un} = \Tr_{I,A} ( (\hat{\pi}_{I}^v \otimes \hat{\pi}_{A}^l \otimes \mathbb{I}) \rho_{tw}) $. In order to normalize, we compute its norm by tracing over its entire Hilbert space [see Der.~\ref{B2} in \ref{appB}]:
\begin{align}
\Tr (\rho_B^{un}) & \approx \Tr (\hat{\pi}_{I}^v \rho_{I}) \Tr ((\hat{\pi}_{A}^l \otimes \mathbb{I}) \rho_{AB}) ( ( 1 \nonumber\\
& - ig  ( \hat{\Pi}_I )_w^* ( \hat{A} )_{w'}^* +  ig ( \hat{\Pi}_I )_w ( \hat{A} )_{w'} )).
\end{align}
Here, we have defined the complex entity $ ( \hat{A} )_{w'}$ to be the \textit{weak-partial-value}. ``Partial'' because while the quantum state (density matrix $ \rho_{AB} $) appearing in it is bipartite and the trace operation is performed over the entire Hilbert space, the system measurement observable $ \hat{A} $ and the post-selection projector $ \pi_A^l $ both act only on part $ A $ of $ \rho_{AB} $\footnote{This can in a restricted sense be also defined as the weak value of $ \hat{A} \otimes \mathbb{I}$
if one defines the projection operator to be $ \pi_A^l \otimes \mathbb{I} $.}. Thus, we have 
\begin{equation}
( \hat{A} )_{w'} \equiv \frac{\Tr ( (\hat{\pi}_{A}^l \hat{A} \otimes \mathbb{I}) \rho_{AB} )}{\Tr ( (\hat{\pi}_{A}^l \otimes \mathbb{I}) \rho_{AB} )} .
\end{equation}
Bob will now measure expectation value of the observable $ \hat{B} $ with respect to the above state (after normalization) using the complex decomposition of the weak-partial-value: $ ( \hat{A} )_{w'} = \Re ( \hat{A} )_{w'} + i \Im ( \hat{A} )_{w'}$ [see Der.~\ref{B5} in \ref{appB}]. Note that the expectation value $ \langle \hat{B} \rangle_f^{Re} $ in the second set of runs would be different from the first set. Since we already know the imaginary part of the weak value from the first set of runs, the real part can be obtained from the second set of runs:
\begin{widetext}
\begin{gather}
\Re ( \hat{\Pi}_I )_w = \nonumber\\
\label{re}
\dfrac{\langle \hat{B} \rangle_f^{Re} \times \Tr ( (\hat{\pi}_{A}^l \otimes \mathbb{I}) \rho_{AB} ) - \Tr ( (\hat{\pi}_{A}^l \otimes \hat{B}) \rho_{AB} ) - g \Im ( \hat{\Pi}_I )_w \bigg( 2 \Re ( \hat{A} )_{w'} \Tr ( (\hat{\pi}_{A}^l \otimes \hat{B}) \rho_{AB} ) + \Tr ( (\hat{\pi}_{A}^l \otimes \hat{B}) \{ (\hat{A} \otimes \mathbb{I}), \rho_{AB} \} ) \bigg)}{ 2 g \Im ( \hat{A} )_{w'} \Tr ( (\hat{\pi}_{A}^l \otimes \hat{B}) \rho_{AB} ) + i g \Tr ( (\hat{\pi}_{A}^l \otimes \hat{B}) [(\hat{A} \otimes \mathbb{I}), \rho_{AB}] ) }. 
\end{gather}
\end{widetext}
Here, $ [x,y] $ and $\{x,y\} $ denote the commutator and the anticommutator respectively.

\subsection{Pre-decided entities}
\label{pde}
In order to obtain the full weak value using expressions \ref{im} and \ref{re}, Bob must know the shared state $ \rho_{AB} $ , the weak interaction observable $ \hat{A} $, the projector $ \hat{\pi}_{A}^l $ used by Alice for the post-selection and the interaction strength $ g $. To allow Bob obtain the full quantum state, they must have also pre-decided the mutually unbiased bases corresponding to the observables involved in the weak interaction and the post-selection. The sequence in which projectors from these basis sets will be used for the weak measurement must be fixed so as to ensure that Bob knows what projector corresponds to the weak value he obtained in a given round of the protocol. Also, they must know the dimensionality of the unknown quantum state so that basis sets with appropriate number of elements can be chosen. These entities can be easily fixed by communicating over a public or encrypted classical channel or by Alice and Bob being at the same location prior to commencement of the protocol since these entities do not change with the quantum state whose information is transferred, as long its dimensionality is constant. Also, Bob's knowledge of the shared state over a noisy channel can be fixed by doing a distance-dependent pre-analysis of the channel noise, losses and decoherence (see Werner state example section~\ref{noise}) before the protocol begins so that Bob will know the shared state even if the communication distance changes. Upon obtaining all the weak values and normalizing, Bob can find the overall factor $ \sqrt{d} \braket{b_0}{\psi} $ to express the full quantum state (ignoring the overall phase).

\subsection{Classical communication requirements}
\label{cit}
\begin{figure}
\centering
\includegraphics[trim=1.5cm 1cm 1cm 1cm,clip=false,angle=0,width=7cm,height=7.5cm]{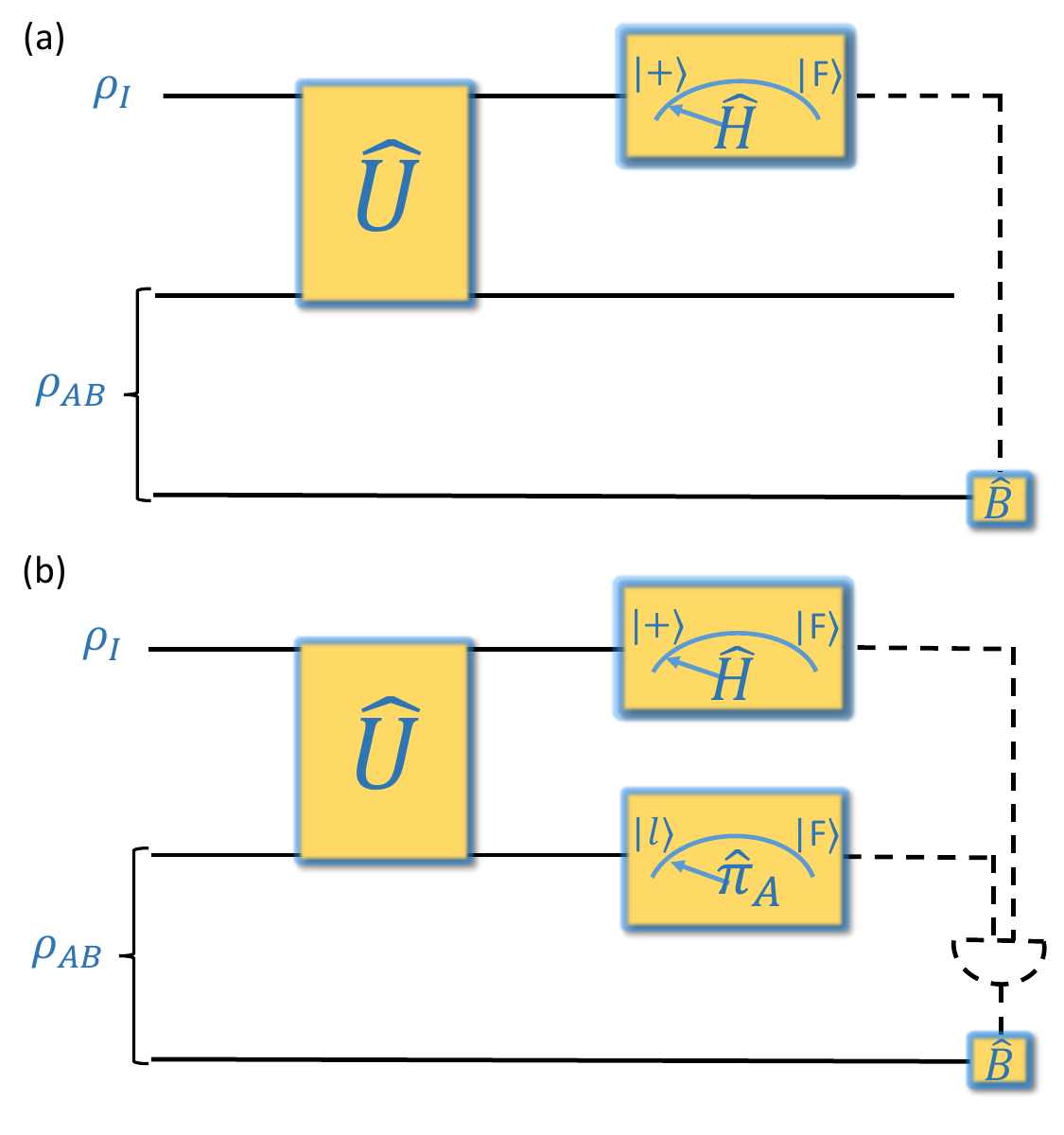}
\caption{(a) The first set of experimental runs entails post-selection only on Alice's system. Bob will measure observable $ \hat{B} $ upon receiving the classical bit $1$. Classical channel is represented by a dashed line. Upon action of the Hadamard operator $ \hat{H} $ on the state of interest, $ \ket{+} $ and $\ket{F}$ represent post-selection success and failure and correspond to $1$ and $0$ on the classical channel respectively. (b) The second set of experimental runs requires post-selection on Alice's system as well as part $A$ of the shared resource. $ \ket{l} $ corresponds to successful post-selection and its outcome is carried by the second classical channel. Bob measures observable $ \hat{B} $ upon receiving the outcome $1$ from the classical \textit{AND} gate which corresponds to instances of simultaneous post-selection success on parts $I$ and $A$. }
\label{fig1}
\end{figure}

Communication from Alice to Bob via classical channel(s) aids the remote determination of the weak value (see Figure \ref{fig1}).
On receiving the appropriate message, Bob measures the value of observable $ \hat{B} $ with respect to his state. After sufficiently many such measurements, the recorded statistics will give him the expectation value of $ \hat{B} $. 
The total number of classical bits communicated from Alice to Bob is given by the sum of bits communicated during the first and second sets of experimental runs corresponding to the real ($ C_{Rk} $) and imaginary ($ C_{Ik} $) parts of each weak value respectively. Thus, we have $ C = \sum_{k = 0}^{d-1} ( C_{Ik} + C_{Rk} ) $:
\begin{equation}
\label{cbits}
C = N \sum_{k = 0}^{d-1} \bigg[ \Tr ( \hat{\pi}_{I}^v \rho_{twk}^2 ) + \Tr ( ( \hat{\pi}_{I}^v \otimes \hat{\pi}_{A}^l \otimes \mathbb{I}) \rho_{twk}^2 ) \bigg],
\end{equation}
where $ N $ is the number of shared copies of the resource in a single set of experimental runs (total number is $ 2Nd $), $d$ is the dimensionality of the unknown quantum state, $ k $ indexes the weak interaction of the $ k^{th} $ projector and the two entities appearing in the parentheses are probabilities of successful post-selection for the $1^{st}$ and $2^{nd}$ set of runs of the protocol respectively. These probabilities are determined by the overlap between the total state after weak interaction and the total state after post-selection. If a continuous variable state is to be transferred, the sum would be over several position eigenkets (or another continuous variable observable) whose weak values are to be remotely measured by Bob to know the full state. After faithfully obtaining a particular expectation value within a certain error threshold (see section on noise), Bob communicates via a reverse classical channel to Alice (total $d$-bits) to proceed to the next weak value. We note that if Alice can choose the mutually unbiased bases such that all the weak values of projectors in which information about the quantum state is encoded are purely imaginary, the second set of experiments would not be necessary. This can happen provided Alice has sufficient information about the quantum state beforehand. Therefore, there is a trade off between complexity of the protocol and the predetermined knowledge of the quantum state whose information is to be transferred.  

\subsection{Proof}	
\label{nsc}
We shall prove that RSD fails, that is, real and imaginary parts of the weak value, as supposed to be obtained by Bob (see Equations.~\ref{im} and \ref{re}), are both equal to zero if and only if $ \rho_{AB} = \rho_{A} \otimes \rho_{B} $. 

We first seek to prove that $ \Im (\hat{\Pi}_I)_w = 0 $ iff $ \rho_{AB} = \rho_{A} \otimes \rho_{B} $. Let us consider set of observables where denominator of $ \Im ( \hat{\Pi}_I )_w $ is not equal to zero as $ \mathcal{B} $. 
Within the set $ \mathcal{B} $, there will always exist an observable basis, that is $ d^2 - 1 $ observables spanning the space of Hermitian observables which is denoted as $ \mathcal{B}_{base} $. Since $ \mathcal{B}_{base} $ is a complete basis, any observable $ \hat{O} $ can be written as linear combination of the members of $ \mathcal{B}_{base} $: $ \hat{O}=\sum_{\hat{B}\in\mathcal{B}_{base}}a_{B}\hat{B} $. $ \langle \hat{B} \rho_{B1} \rangle = \langle \hat{B} \rho_{B2} \rangle $ for all members of $ \mathcal{B}_{base} $ implies that $ \langle \hat{O} \rho_{B1} \rangle = \langle \hat{O} \rho_{B2} \rangle $ which is equivalent to $ \rho_{B1} = \rho_{B2} $.
We consider $ \Im ( \hat{\Pi}_I )_w = \langle \hat{B} \rangle^{Im}_f - \Tr(\hat{B} \rho_B^{in}) = 0 $ while the denominator of Eq.\ref{im} is non-zero 
(see also points (1) and (2) in Appendix.\ref{appC}). This is equivalent to $ \langle \hat{B} \rangle^{Im}_f = \Tr(\hat{B} \rho_B^{in}) = \langle \hat{B} \rangle_{in} $, which is further equivalent to $ \rho_B^f = \rho_B^{in} $ provided the set of observables $ \hat{B} $ -- which must be within the set of observables where denominator is non-zero --
are $d^2 - 1$ spin-1/2 (with their d-dimensional versions) observables in respective higher dimensions (see point (3) in Appendix.\ref{appC}).
This implies $ \rho_{AB} = \rho_{A} \otimes \rho_{B} $.

Furthermore, $ \rho_{AB} = \rho_{A} \otimes \rho_{B} $ implies $ \rho_{B}^N = \rho_{B}^{un}/\Tr(\rho_B^{un}) = \rho_B $ (see Der.~\ref{C1} in Appendix.~\ref{appC}) which is equivalent to $ \Im ( \hat{\Pi}_I )_w = 0 $. Therefore, $ \Im ( \hat{\Pi}_I )_w = 0 $ if and only if $ \rho_{AB} = \rho_{A} \otimes \rho_{B} $. 

By considering $ \Re ( \hat{\Pi}_I )_w = 0 $, we find (see Der.\ref{reprf1}, \ref{reprf2}, \ref{reprf3}, and \ref{reprf4} in Appendix.~\ref{appC}),
\begin{eqnarray}
\langle \hat{B} \rangle^{Re}_f &=& \Tr(\hat{B} \rho_B^{in}) + g \frac{\langle \hat{B} \rangle_f^{Im} - \Tr ( \hat{B} \rho_B^{in} )}{Denominator} \sum_{i=1}^p [T1_l] E_l \nonumber\\
&&+ g \frac{\langle \hat{B} \rangle_f^{Im} - \Tr ( \hat{B} \rho_B^{in} )}{Denominator} \sum_{i=1}^p [T2_l] N_l,
\end{eqnarray}
where $T1_l$ and $T2_l$ are the respective terms in Equation.~\ref{reprf2}. Since $ [\hat{\pi}^l_A,\hat{A}] \neq 0 $, let  $ [\hat{\pi}^l_A ,\hat{A}] = K \implies \hat{\pi}^l_A \hat{A} = K + \hat{A} \hat{\pi}^l_A $. Therefore, $ \{\hat{\pi}^l_A, \hat{A}\} = \hat{\pi}^l_A \hat{A} + \hat{A} \hat{\pi}^l_A = K + 2 \hat{A} \hat{\pi}^l_A \neq 0 $, where $ K \neq -2 \hat{A} \hat{\pi}^l_A $. We also have $ \rho_{AB} = \rho_{AB}^\dagger $. Therefore, $ T1_l \neq 0 $ and $ T2_l \neq 0 $. This implies $ \langle \hat{B} \rangle^{Re}_f = \Tr(\hat{B} \rho^{in}_B) $ if and only if $ \langle \hat{B} \rangle^{Im}_f = \Tr(\hat{B} \rho^{in}_B) $, that is $ \Im(\hat{\Pi}_w) = 0 $. But we have proven that $ \Im(\hat{\Pi}_I)_w = 0 $ if and only if $ \rho_{AB} = \rho_{A} \otimes \rho_{B} $. Hence, $ \langle \hat{B} \rangle^{Re}_f = \Tr(\hat{B} \rho^{in}_B) $, that is, $ \Re(\hat{\Pi}_I)_w = 0 $ if and only if $ \rho_{AB} = \rho_{A} \otimes \rho_{B} $.  

\subsection{Security}
\label{sec}
To demonstrate that our protocol is secure against a classical eavesdropper Eve, consider the following attack: interference in the classical communication channel which transfers bits from the post-selection results to Bob. As is evident from the expression for $C$ (Equation~\ref{cbits}: the number of cbits communicated), these classical bits carry no information about the quantum state that was encoded on Alice's system state, which makes the protocol secure against classical eavesdropping. Another way Eve can interfere is by blocking Bob's part of the entangled/non-product photons or another bipartite resource (for example, a continuous variable resource) that is shared by Alice and Bob. In this kind of attack, Eve can make her own measurements of an observable $\hat{E}$ and get access to the classical communication channel in an attempt to reconstruct Alice's encoded state. This, however, will inform Bob who will notice that he is no longer receiving a part of the bipartite resource states he is supposed to and can inform Alice via a reverse classical channel to stop the communication. Thus, statistical nature of the protocol prevents this kind of eavesdropping.

This security can be viewed in the context of a classical protocol where Alice performs local quantum state tomography and transmits the results to Bob via classical communication. This implementation is not quantum mechanically secure because results of the tomography can only be secured by classical encryption, key distribution, one-time pad methods etc. Eve can hack the encryption system used or directly read off the classical channel if there is no encryption. As such, it does not provide us security beyond what is classically possible. This, in fact, has been the main motivation behind quantum communication protocols such as quantum key distribution and quantum teleportation.

\section{Example}	
\label{example}
We shall demonstrate RSD using the Bell-diagonal state in $ \mathcal{H}^2 \otimes \mathcal{H}^2 $ as the shared resource:
$
	\rho_{AB}=(1/4)[ \mathbb{I} + \sum_{i=1}^{3} c_i (\sigma_{iA} \otimes \sigma_{iB}) ]
$.
Here $ \sigma_i $ represent Pauli matrices in $ \mathcal{H}^2 $ and $c_i \in [-1,1]$.
We choose $ \hat{A} = \begin{bmatrix}
	a_{11} & a_{12} \\
	a_{12}* & a_{22} 
\end{bmatrix} $, $ \hat{B} = \begin{bmatrix}
b_{11} & b_{12} \\
b_{12}* & b_{22} 
\end{bmatrix} $, 
and $ \hat{\pi}_A^l = \ket{\sigma_{zA} = -1}\bra{\sigma_{zA} = -1} $. From these choices, Equations \ref{im} and \ref{re} reduce to simple forms [see Eq.~\ref{D1} and \ref{D2} in \ref{appD}]. Let us consider the transfer of a d-dimensional pure state $ \ket{\psi} = \sum_{k=0}^{d-1} a_k \ket{a_k} $. Therefore, $ \ket{b_0} = (1/\sqrt{d}) \sum_{k=0}^{d-1} \ket{a_k} $ and all the weak values would be $ ( \hat{\Pi}_I^m )_w = a_{m} / (\sum_{k=0}^{d-1} a_k ) $. Plugging these entities in Eq.~\ref{cbits}, using $ \hat{\pi}^v_I = \ket{b_0}\bra{b_0} $, and $ \rho_{I} = \ket{\psi}\bra{\psi} $, we have
$
C = (3Nd/2) \Tr(\hat{\pi}^v_I \rho_{I}) \Tr(\rho_{AB}^2)
$ (see Eq.~\ref{D5} in Appendix~\ref{appD}), where $\Tr(\rho_{AB}^2) = 1/4(1 + c_1^2 + c_2^2 + c_3^3)$ is the purity of the Bell-diagonal state. This reflects the probabilities during first and second set of experimental runs, given by $ d \Tr ( \hat{\pi}_{I}^v \rho_I ) \Tr ( \rho_{AB}^2 ) $ and $ (d/2) \Tr ( \hat{\pi}_{I}^v \rho_I ) \Tr ( \rho_{AB}^2 ) $  respectively (see Eq.~\ref{D4} in Appendix~\ref{appD}).

\section{Remarks on noise, error and experimental implementation}
\label{noise}
Statistical error in determining the state $ \ket{\psi}_I $ on Bob's side originates from his measurement of the expected value of $ \hat{B} $ and propagates~\cite{ku1966notes} through the expressions \ref{im} and \ref{re} for the real and imaginary parts of the individual weak values respectively. In Eq.~\ref{im}, replacing $\Im ( \hat{\Pi}_I^k )_w \equiv W_{Ik}$, $ \langle \hat{B} \rangle_f^{Im} \equiv B_{Ik} $, $ \Tr ( \hat{B} \rho_B^{in} ) \equiv A $ and the denominator with $ X_{Ik} $, we have,
\begin{equation}
\label{imer}
W_{Ik} = \frac{B_{Ik} - A}{X_{Ik}}.
\end{equation}
Thus, we get the error to be,
\begin{eqnarray*}
\label{deltai}
(\Delta W_{Ik})^2 = \bigg( \frac{\partial W_{Ik}}{\partial B_{Ik}} \bigg)^2 (\Delta B_{Ik})^2
= \bigg( \frac{\Delta B_{Ik}}{X_{Ik}} \bigg)^2.
\end{eqnarray*}
Likewise, in Eq.~\ref{re}, replacing $\Re ( \hat{\Pi}_I^k )_w \equiv W_{Rk} $, $ \langle \hat{B} \rangle_f^{Re} \equiv B_{Rk} $, the other constant elements in the numerator and denominator as $ Y_{1Rk} $, $ Y_{2Rk} $, $ Y_{3Rk} $ and $ Y_{4Rk} $ respectively, 
and using Eq.~\ref{imer}, we get
\begin{eqnarray*}
\label{reer}
W_{Rk} &=& \frac{X_{Ik} B_{Rk} Y_{1Rk} - Y_{2Rk} X_{Ik} - B_{Ik} Y_{3Rk} + A Y_{3Rk}}{Y_{4Rk} X_{Ik}}.
\end{eqnarray*} 
Thus, the error is,
\begin{eqnarray*}
\label{deltar}
(\Delta W_{Rk})^2 &=& \bigg( \frac{\partial W_{Rk}}{\partial B_{Rk}} \bigg)^2 (\Delta B_{Rk})^2
+ \bigg( \frac{\partial W_{Rk}}{\partial B_{Ik}} \bigg)^2 (\Delta B_{Ik})^2 \nonumber\\
&=& \bigg( \frac{Y_{1Rk} \Delta B_{Rk}}{Y_{4Rk}} \bigg)^2 
+ \bigg( \frac{Y_{3Rk} \Delta B_{Ik}}{Y_{4R} X_{Ik}} \bigg)^2.
\end{eqnarray*}
Considering the standard error scaling~\cite{ku1966notes} when obtaining the expectation values $\Delta B_{Rk}$ and $ \Delta B_{Ik} $ as proportional to $ 1/\sqrt{C_{Rk}} $ and $ 1/\sqrt{C_{Ik}} $ respectively. Combining this with the above, we see that errors in determining respective weak values and in effect, coefficients characterizing the state scale as $\sim 1/\sqrt{C} $ and therefore as $ \sim 1/\sqrt{\Tr (\rho_{AB}^2)} $ [see Der.~\ref{D3} in \ref{appD}]. 

It is known that sharing a resource state over a noisy quantum channel decreases its purity. A tractable example to demonstrate this would be the Werner state, with a singlet content quantified by $z \in [0,1] $, and a non-product nature for the full range of $z$ encompassing the regimes of discord and entanglement: $ \rho_w$ $=$ $z \ket{\psi^-}\bra{\psi^-} + \frac{1-z}{4} \mathbb{I}_4 $. Its purity is given by $ (1/4) ( 1 + 3 z^2 ) $. Upon sending this state through a decoherent optical fibre~\cite{2016APS..DMP.H4009G,MKGThesis16}, it changes to
\begin{equation}
\rho_w' = \left(
\begin{array}{cccc}
\frac{1-z}{4} & 0 & 0 & 0 \\
0 & \frac{z+1}{4} & -\frac{z}{2} e^{-4 \Delta \phi^2} & 0 \\
0 & -\frac{z}{2} e^{-4 \Delta \phi^2} & \frac{z+1}{4} & 0 \\
0 & 0 & 0 & \frac{1-z}{4} \\
\end{array}
\right).
\end{equation}
The purity now becomes $ (1/4) [ 1 + (1 + 2 e^{-8 \Delta \phi ^2} ) z^2 ] $ clearly indicating $ \Tr (\rho_{w}'^2) < \Tr (\rho_w^2) $. Suppose the accepted error threshold for a faithful run of the protocol demands $N$ copies of the Werner state shared via noiseless channels. To obtain the same efficiency when the resources are shared over noisy channels, one would have to increase the number of copies shared to $ N' $ given by
\begin{eqnarray}
N' &=& \ceil{\frac{N \Tr (\rho_{w}^2)}{\Tr (\rho_{w}'^2)}}. 
\end{eqnarray}
This would also allow to switch between fibers with different noise profiles or different lengths without compromising the faithfulness of the protocol. For example, typical range of $ \Delta \phi $ corresponding to telecommunication-wavelength for an optical fiber of 500 meters is 190-250 radians~\cite{2016APS..DMP.H4009G,MKGThesis16}. Comparing the case where there is no noise ($N$ shared copies) to the case where $ \Delta \phi = 200 \text{rad} $ ($N'$ shared copies) for a state characterized by $ z = 0.4 $ would give us $ N' = 1.27586 N $. Thus, irrespective of the amount of noise in the channel, faithfulness of the protocol would not be compromised provided enough number of noisy but non-product resource states are shared by the two parties. It may also be noted that the protocol would continue to work even if $ z < 1/3 $, when the Werner state is solely discordant~\cite{modi2012classical,pande2017minimum}. Here the sole focus was on the error due to noise in the channel reflected statistically through the entity $ \langle \hat{B} \rangle_f^{Re/Im} $. A full reconstruction of the state possible either through real~\cite{PhysRevLett.117.120401} or simulated experiments~\cite{maccone2014state} would allow a comparison between the state determined by Bob and the state that was intended to be transferred using a metric such as fidelity or trace distance. This would also account for the error pertaining to imperfectly restricting the coupling interaction strength to the first order due to the weak approximation.

Due to the flexibility in choice of the resource state with regard to dimensionality as well as purity, the protocol can, in principle, be implemented in all architectures~\cite{yonezawa2004demonstration,nielsen1998complete,PhysRevLett.118.160501,xu2010experimental} that admit at least one kind of non-product resources -- whether shared Bell pairs, Laguerre Gauss (LG)-mode pointer states with non-zero orbital angular momentum (OAM), or entangled multi-mode Gaussian states, among others. We also note that, since multiple non-product states are necessary for the protocol to work, we want an experimental setup which can produce bipartite entangled (a subset of non-product) states with a consistently high rate. This also opens the tantalizing possibility of implementing the protocol by constructing quantum communication networks involving more than two parties~\cite{bae2005three,pirandola2005conditioning,pirandola2015advances,yonezawa2004demonstration,hu2016experimental}, even when it is difficult to maintain a high degree of multipartite entanglement. 

\section{Conclusion and Outlook}
\label{conclusion}
In essence, we have developed a method to transfer information of an unknown quantum state of any known dimensions, encompassing continuous variable states, from one party to another spatially separated party using a non-product bipartite quantum state of any dimensionality as a resource. The fundamental principle underlying RSD as well as other quantum communication protocols like teleportation and remote state preparation~\cite{bennett1993teleporting,pati2000minimum,lo2000classical,bennett2001remote,gisin2007quantum} is the creation of \textit{transitive correlation} between parts $ I $ and $ B $ due to the von Neumann interaction~\cite{von1955mathematical} and the subsequent entanglement caused between $ I $ and $ A $ and the correlation (encompassing non-product nature of all states in our case) which is already present between $ A $ and $ B $. In case of teleportation, the von Neumann interaction is strong and translates to the \textit{C-NOT} gate for qubit registers~\cite{nielsen2010quantum}. For remote state preparation, the strong interaction translates to a \textit{C-U} gate (controlled unitary~\cite{nielsen2010quantum}), where the rotation $U$ is determined by Alice's knowledge of the quantum state which is to be prepared at Bob's end. The transitive correlation facilitates information sharing between spatially separated parts. E-QKD uses direct correlation by sending photons of an entangled state to Alice and Bob and it is known that an entanglement-breaking channel does not allow M-QKD to succeed~\cite{scarani2009the}. In this protocol, we operationally exploit all correlations manifested in any non-product state~\cite{roszak2015relation,wiseman2007steering,buscemi2012all,guo2014quantum,masanes2008all,peres1996separability,maccone2015complementarity,PhysRevLett.104.140404,PhysRevA.73.022311,PhysRevA.81.052318} using local operations and classical communication. In light of this, it may be interesting to pursue robust non-locality criteria~\cite{bell1964einstein,roszak2015relation,werner1989quantum,wiseman2007steering,masanes2008all,buscemi2012all,silva2015multiple,gisin1991bell} which encompass such wide class of correlations. 

It is pertinent to note that quantum correlations beyond entanglement -- specifically quantum discord -- have been used in the deterministic quantum computation with one pure qubit (DQC1) model~\cite{lanyon2008experimental} as well as in a slight variant of quantum teleportation~\cite{wang2012quantum} before. In the latter method, information about a qubit state can be transmitted using a mixed two-qubit resource state which may have no entanglement, and should be followed by state tomography on Bob's reduced density matrix to determine parameters of the state that was meant to be transmitted by Alice. So, the limitations of teleportation get carried over to this method (also see below). 

Conventional analogues of the current protocol would be (i) teleportation followed by state tomography~\cite{vogel1989determination,cramer2010efficient}, and (ii) tomography followed by E-QKD or M-QKD where the implementation challenges will apply when transmitting amplitudes of the state. Specifically, on lines similar to Ref.~\cite{maccone2014state}, one may be able to perform an analysis comparing this protocol to (i), provided the exponential scaling with dimensionality of the state at hand in terms of the number of measurements for specific sets of observables to determine the respective probabilities and phases can be managed efficiently -- a difficult task. Such an analysis, of course, would be possible only if a given resource state can be used to teleport the state of interest. For the many state-resource pairs where this is not possible, the protocol could serve as an alternative method for quantum state transfer provided a reliable state preparation mechanism is in place at the receiver's end, thus enabling high-dimensional distributed quantum computing. Another point of comparison with QKD is the determination of data transmission rates possible with the protocol and its variants. While the protocol is trivially secure against attack by a classical eavesdropper (assuming he even knows the basis), it would be interesting to investigate security against an attack by a quantum eavesdropper as done for teleportation~\cite{PhysRevA.64.010301} and quantum key distribution~\cite{lo2014secure}. These investigations to operationalize the protocol further will be taken up in future work.

In addition to quantum information transfer, remote determination of a single weak value in itself is useful in that all of the characteristics of the weak value and the weak measurement are now available to be explored remotely. These include parameter estimation via weak value amplification~\cite{coto2017power}, resolution of quantum paradoxes etc.~[see Ref.~\cite{dressel2014colloquium} for an instructive review]. During development of the protocol, we introduced an entity called the weak-partial-value, where the interaction and post-selection is performed only on a part of a non-product state. It can be generalized to an entire class wherein Hilbert space selective weak interaction(s) and post-selection(s) are performed. Such quantities might indeed arise when the protocol is extended to enable communication between more than two parties. It is therefore worthwhile investigating the properties, their implications, and operational significance of the weak-partial-value. The protocol could be extended to enable information transfer of mixed states if one is able to remotely determine the joint weak values and use these to express the joint weak averages which constitute all elements of the density matrix~\cite{lundeen2012procedure}, and it could be implemented using certain experimental methods~\cite{mirhosseini2014compressive,PhysRevLett.127.040402} to achieve remote determination of high-dimensional states. Although it is difficult to achieve the former with a general pointer state, it could be possible with the non-product tripartite version of specific pointers like separable (on one part) Gaussian~\cite{PhysRevLett.92.130402} or the LG-mode pointer states~\cite{PhysRevLett.109.040401} with non-zero OAM. 
	
\textit{Note added.---} After completion of this manuscript, 
a slightly related work by Ref.~\cite{pati2013weak} came to notice. There, the post-selection is performed by Bob, dimensions of system and resource state are interdependent, and the method depends on entanglement of the resource state.

\subsection*{Acknowledgements}
I am grateful to Charles H.~Bennett, Raul Coto, Justin Dressel, Yuji Hasegawa, Dipankar Home, T.~S.~Mahesh, Jonathan Oppenheim, Arun Kumar Pati, Ujjwal Sen, Anil Shaji, Lev Vaidman, and Giuseppe Vallone for useful feedback. I thank Manish Gupta for discussions on noise in optical fibers, and Soumik Adhikary, Jacob Biamonte, Daniil Rabinovich, Richik Sengupta, and Akshay Vishwanathan for pointing out that the proof in an earlier version was incomplete. Posters on this work were presented at the ``3rd International Conference on Quantum Foundations" at NIT Patna, at the ``International Symposium on New Frontiers in Quantum Correlations" at SNBNCBS, Kolkata, and at ``Quantum Optics X" at Toruń, Poland. V.~R.~P. acknowledges support from the DST, Govt.~of India through the INSPIRE scheme and the hospitality of Bose Institute. Research at HRI is supported by the DAE, Govt.~of India.

\subsection*{Data availability}
This manuscript has no available data.

\appendix

\onecolumngrid

\section{Protocol algebraic details -- transferring imaginary part of the weak value}
\label{appA}
\begin{enumerate}
	\item Bob's unnormalized state corresponding to the measurement of the imaginary part of the weak value:
	\begin{eqnarray}
	\label{A1}
	\rho_B^{un} &=& \Tr_{I,A} ( \hat{\pi}_{I}^v \rho_{tw}) \nonumber\\
	&\approx& \Tr_{I,A} (\hat{\pi}_{I}^v (\mathbb{I} + ig\hat{\Pi}_I \otimes \hat{A} \otimes \mathbb{I}) \rho_{I} \otimes \rho_{AB} (\mathbb{I} - ig\hat{\Pi}_I \otimes \hat{A} \otimes \mathbb{I})) \nonumber\\
	&=& \Tr_{I,A} (\hat{\pi}_{I}^v (\rho_{I} \otimes \rho_{AB} + ig\hat{\Pi}_I \rho_{I} \otimes (\hat{A} \otimes \mathbb{I}) \rho_{AB} ) (\mathbb{I} - ig\hat{\Pi}_I \otimes \hat{A} \otimes \mathbb{I})) \nonumber\\
	&=& \Tr_{I,A} (\hat{\pi}_{I}^v (\rho_{I} \otimes \rho_{AB}  - ig \rho_{I} \hat{\Pi}_I \otimes \rho_{AB} ( \hat{A} \otimes \mathbb{I}) + ig\hat{\Pi}_I \rho_{I} \otimes (\hat{A} \otimes \mathbb{I}) \rho_{AB} ) ) \nonumber\\
	&=& \Tr_{I,A} ( (\hat{\pi}_{I}^v \rho_{I} \otimes \rho_{AB})  - ig \hat{\pi}_{I}^v \rho_{I} \hat{\Pi}_I \otimes \rho_{AB} ( \hat{A} \otimes \mathbb{I}) + ig ( \hat{\pi}_{I}^v \hat{\Pi}_I \rho_{I} ) \otimes (\hat{A} \otimes \mathbb{I}) \rho_{AB} )  \nonumber\\
	&=& \Tr_{I} (\hat{\pi}_{I}^v \rho_{I}) ( ( \Tr_{A} ( \rho_{AB})  - ig  ( \hat{\Pi}_I )_w^* \Tr_{A} ( \rho_{AB} ( \hat{A} \otimes \mathbb{I})) + ig ( \hat{\Pi}_I )_w \Tr_{A} ( (\hat{A} \otimes \mathbb{I}) \rho_{AB} ) )).
	\end{eqnarray} 
	\item Normalizing the above state: \\
	Cyclic property of the trace allows us to write $ \Tr_{A} ( \rho_{AB} ( \hat{A} \otimes \mathbb{I})) = \Tr_{A} ( (\hat{A} \otimes \mathbb{I}) \rho_{AB} ) $. Bob's initial state is $ \Tr_{A} ( \rho_{AB}) \equiv \rho_B^{in} $. Thus, we can write
	\begin{eqnarray*}
		\rho_B^{un} &\approx& \Tr_{I} (\hat{\pi}_{I}^v \rho_{I}) (\rho_B^{in} - 2 g \Im ( \hat{\Pi}_I )_w \Tr_{A} ((\hat{A} \otimes \mathbb{I}) \rho_{AB}) ).
	\end{eqnarray*} 
	Further, using $ \Tr (\rho_B^{in}) = 1 $:
	\begin{eqnarray}
	\label{A2}
	\rho_{B}^N = \frac{\rho_{B}^{un}}{\Tr(\rho_B^{un})} &\approx& \frac{\bigg(\rho_B^{in} - 2 g \Im ( \hat{\Pi}_I )_w \Tr_{A} ((\hat{A} \otimes \mathbb{I}) \rho_{AB}) \bigg)}{\bigg( 1 - 2 g \Im ( \hat{\Pi}_I )_w \Tr ( (\hat{A} \otimes \mathbb{I}) \rho_{AB} ) \bigg)}. \nonumber\\
	\end{eqnarray}
	In line with the weak approximation, one can bring the denominator to the numerator and Taylor expand up to the first order in $g$:
	\begin{eqnarray}
	\label{A3}
	\rho_B^N &\approx& \bigg( \rho_B^{in} - 2 g \Im ( \hat{\Pi}_I )_w \Tr_{A} ((\hat{A} \otimes \mathbb{I}) \rho_{AB}) \bigg) \bigg(1 + 2 g \Im ( \hat{\Pi}_I )_w \Tr ((\hat{A} \otimes \mathbb{I}) \rho_{AB}) \bigg)  \nonumber\\
	&=& \rho_B^{in} + 2g \Im ( \hat{\Pi}_I )_w \Tr((\hat{A} \otimes \mathbb{I})\rho_{AB}) \rho_{B}^{in} - 2g \Im ( \hat{\Pi}_I )_w \Tr_A ((\hat{A} \otimes \mathbb{I})\rho_{AB}) \nonumber\\
	&=& \rho_B^{in} + 2 g \Im ( \hat{\Pi}_I )_w \bigg(\Tr ((\hat{A} \otimes  \mathbb{I})\rho_{AB})\rho_{B}^{in} - \Tr_{A} ((\hat{A} \otimes \mathbb{I})\rho_{AB})\bigg). \nonumber\\
	\end{eqnarray}
	\item Bob's expectation value:
	\begin{eqnarray}
	\label{A4}
	\langle \hat{B} \rangle_f^{Im} = \Tr (\hat{B} \rho_B^N)
	\approx \Tr(\hat{B} \rho_B^{in}) + 2 g \Im ( \hat{\Pi}_I )_w \bigg( \Tr ((\hat{A}\otimes \mathbb{I})\rho_{AB}) \Tr(\hat{B}\rho_B^{in}) - \Tr(\hat{B} \Tr_{A} ((\hat{A} \otimes \mathbb{I})\rho_{AB}))\bigg).
	\end{eqnarray}
\end{enumerate}

\section{Protocol algebraic details -- transferring real part of the weak value}
\label{appB}
\begin{enumerate}
	\item Bob's unnormalized state:
	\begin{eqnarray*}
		\rho_B^{un} &=& \Tr_{I,A} ((\hat{\pi}_{I}^v \otimes \hat{\pi}_{A}^l \otimes \mathbb{I}) \rho_{tw}) \nonumber\\
		&\approx& \Tr_{I,A} ((\hat{\pi}_{I}^v \otimes \hat{\pi}_{A}^l \otimes \mathbb{I}) (\mathbb{I} + ig\hat{\Pi}_I \otimes \hat{A} \otimes \mathbb{I}) \rho_{I} \otimes \rho_{AB} (\mathbb{I} - ig\hat{\Pi}_I \otimes \hat{A} \otimes \mathbb{I})) \nonumber\\
		&=& \Tr_{I,A} ((\hat{\pi}_{I}^v \otimes \hat{\pi}_{A}^l \otimes \mathbb{I}) (\rho_{I} \otimes \rho_{AB} + ig\hat{\Pi}_I \rho_{I} \otimes (\hat{A} \otimes \mathbb{I}) \rho_{AB} ) (\mathbb{I} - ig\hat{\Pi}_I \otimes \hat{A} \otimes \mathbb{I})) \nonumber\\
		&=& \Tr_{I,A} ((\hat{\pi}_{I}^v \otimes \hat{\pi}_{A}^l \otimes \mathbb{I}) (\rho_{I} \otimes \rho_{AB}  - ig \rho_{I} \hat{\Pi}_I \otimes \rho_{AB} ( \hat{A} \otimes \mathbb{I}) + ig\hat{\Pi}_I \rho_{I} \otimes (\hat{A} \otimes \mathbb{I}) \rho_{AB} ) ) \nonumber\\
		&=& \Tr_{I,A} ( (\hat{\pi}_{I}^v \rho_{I} \otimes (\hat{\pi}_{A}^l \otimes \mathbb{I}) \rho_{AB}  - ig \hat{\pi}_{I}^v \rho_{I} \hat{\Pi}_I \otimes (\hat{\pi}_{A}^l \otimes \mathbb{I}) \rho_{AB} ( \hat{A} \otimes \mathbb{I}) + ig \hat{\pi}_{I}^v \hat{\Pi}_I \rho_{I} \otimes (\hat{\pi}_{A}^l \otimes \mathbb{I}) (\hat{A} \otimes \mathbb{I}) \rho_{AB} ) ) \nonumber\\
		&=& \Tr_{I} (\hat{\pi}_{I}^v \rho_{I})  ( \Tr_{A} ((\hat{\pi}_{A}^l \otimes \mathbb{I}) \rho_{AB})  - ig  ( \hat{\Pi}_I )_w^* \Tr_{A} ( (\hat{\pi}_{A}^l \otimes \mathbb{I}) \rho_{AB} ( \hat{A} \otimes \mathbb{I})) + ig ( \hat{\Pi}_I )_w \Tr_{A} ( (\hat{\pi}_{A}^l \otimes \mathbb{I}) (\hat{A} \otimes \mathbb{I}) \rho_{AB} ) ).
	\end{eqnarray*}
	Taking the partial trace operation over $ I $ and $ A $ inside the parenthesis, one finds:
	\begin{eqnarray}
	\label{B1}
	\rho_B^{un} &\approx& ( ( \Tr_{I} (\hat{\pi}_{I}^v \rho_{I}) \Tr_{A} ((\hat{\pi}_{A}^l \otimes \mathbb{I}) \rho_{AB}) - ig \Tr_{I} ( \hat{\pi}_{I}^v \rho_{I} \hat{\Pi}_I) \Tr_{A}( (\hat{\pi}_{A}^l \otimes \mathbb{I}) \rho_{AB} ( \hat{A} \otimes \mathbb{I})) + ig \Tr_{I} ( \hat{\pi}_{I}^v \hat{\Pi}_I \rho_{I}) \Tr_{A} ( (\hat{\pi}_{A}^l \otimes \mathbb{I}) (\hat{A} \otimes \mathbb{I}) \rho_{AB} ) )) \nonumber\\
	&=& \Tr_{I} (\hat{\pi}_{I}^v \rho_{I}) ( ( \Tr_{A} ((\hat{\pi}_{A}^l \otimes \mathbb{I}) \rho_{AB}) - ig \Tr_{I} ( \hat{\pi}_{I}^v \rho_{I} \hat{\Pi}_I) \Tr_{A}( (\hat{\pi}_{A}^l \otimes \mathbb{I}) \rho_{AB} ( \hat{A} \otimes \mathbb{I})) + ig \Tr_{I} ( \hat{\pi}_{I}^v \hat{\Pi}_I \rho_{I}) \Tr_{A} ( (\hat{\pi}_{A}^l \otimes \mathbb{I}) (\hat{A} \otimes \mathbb{I}) \rho_{AB} ) )) \nonumber\\
	&=& \Tr_{I} (\hat{\pi}_{I}^v \rho_{I}) ( \Tr_{A} ((\hat{\pi}_{A}^l \otimes \mathbb{I}) \rho_{AB}) - ig  ( \hat{\Pi}_I )_w^* \Tr_{A} ( (\hat{\pi}_{A}^l \otimes \mathbb{I}) \rho_{AB} ( \hat{A} \otimes \mathbb{I})) + ig ( \hat{\Pi}_I )_w \Tr_{A} ( (\hat{\pi}_{A}^l \otimes \mathbb{I}) (\hat{A} \otimes \mathbb{I}) \rho_{AB} ) ).
	\end{eqnarray}
	\item Trace of the unnormalized state:
	\begin{eqnarray}
	\label{B2}
	\Tr (\rho_b^{un}) &\approx& \Tr ( \Tr_{I} (\hat{\pi}_{I}^v \rho_{I}) ( ( \Tr_{A} ((\hat{\pi}_{A}^l \otimes \mathbb{I}) \rho_{AB}) - ig  ( \hat{\Pi}_I )_w^* \Tr_{A}( (\hat{\pi}_{A}^l \otimes \mathbb{I}) \rho_{AB} ( \hat{A} \otimes \mathbb{I})) + ig ( \hat{\Pi}_I )_w \Tr ( (\hat{\pi}_{A}^l \otimes \mathbb{I}) (\hat{A} \otimes \mathbb{I}) \rho_{AB} ) ))) \nonumber\\
	&=&  \Tr_{I} (\hat{\pi}_{I}^v \rho_{I}) ( ( \Tr ((\hat{\pi}_{A}^l \otimes \mathbb{I}) \rho_{AB}) - ig  ( \hat{\Pi}_I )_w^* \Tr( (\hat{\pi}_{A}^l \otimes \mathbb{I}) \rho_{AB} ( \hat{A} \otimes \mathbb{I})) +  ig ( \hat{\Pi}_I )_w \Tr ( (\hat{\pi}_{A}^l \otimes \mathbb{I}) (\hat{A} \otimes \mathbb{I}) \rho_{AB} ) ))\nonumber\\ 
	&=& \Tr_{I} (\hat{\pi}_{I}^v \rho_{I}) \Tr ((\hat{\pi}_{A}^l \otimes \mathbb{I}) \rho_{AB}) ( ( 1 - ig  ( \hat{\Pi}_I )_w^* ( \hat{A} )_{w'}^* + ig ( \hat{\Pi}_I )_w ( \hat{A} )_{w'} )). 
	\end{eqnarray}
	\item Now, let us normalize Bob's state using its norm in the denominator:
	\begin{gather}
	\label{B3}
	\rho_B^{N} = \frac{\rho_B^{un}}{\Tr (\rho_B^{un})} \nonumber\\
	\approx \frac{ \Tr_{I} (\hat{\pi}_{I}^v \rho_{I}) ( ( \Tr_{A} ((\hat{\pi}_{A}^l \otimes \mathbb{I}) \rho_{AB}) - ig  ( \hat{\Pi}_I )_w^* \Tr_{A}( (\hat{\pi}_{A}^l \otimes \mathbb{I})\rho_{AB} ( \hat{A} \otimes \mathbb{I})) + ig ( \hat{\Pi}_I )_w \Tr_{A} ( (\hat{\pi}_{A}^l \otimes \mathbb{I}) (\hat{A} \otimes \mathbb{I}) \rho_{AB} ) }{ \Tr_{I} (\hat{\pi}_{I}^v \rho_{I}) \Tr ((\hat{\pi}_{A}^l \otimes \mathbb{I}) \rho_{AB}) (  1 - ig  ( \hat{\Pi}_I )_w^* ( \hat{A} )_{w'}^* + ig ( \hat{\Pi}_I )_w ( \hat{A} )_{w'} )}. 
	\end{gather}
	Using the weak approximation, the inverse of the denominator can be Taylor expanded up to the first order in $g$:
	\begin{eqnarray}
	\label{B4}
	\rho_B^{N} &\approx& \frac{1}{\Tr ((\hat{\pi}_{A}^l \otimes \mathbb{I}) \rho_{AB})} (  \Tr_{A} ((\hat{\pi}_{A}^l \otimes \mathbb{I}) \rho_{AB})  - ig  ( \hat{\Pi}_I )_w^* \Tr_{A}( (\hat{\pi}_{A}^l \otimes \mathbb{I}) \rho_{AB} ( \hat{A} \otimes \mathbb{I})) \nonumber\\
	&&+ ig ( \hat{\Pi}_I )_w \Tr_{A} ( (\hat{\pi}_{A}^l \otimes \mathbb{I}) (\hat{A} \otimes \mathbb{I}) \rho_{AB} )) ( 1 + ig  ( \hat{\Pi}_I )_w^* ( \hat{A} )_{w'}^* - ig ( \hat{\Pi}_I )_w ( \hat{A} )_{w'} )  \nonumber\\
	&=& \frac{1}{\Tr ((\hat{\pi}_{A}^l \otimes \mathbb{I}) \rho_{AB})} (  \Tr_{A} ((\hat{\pi}_{A}^l \otimes \mathbb{I}) \rho_{AB}) + ig  ( \hat{\Pi}_I )_w^* ( \hat{A} )_{w'}^* \Tr_{A} ((\hat{\pi}_{A}^l \otimes \mathbb{I}) \rho_{AB}) 
	- ig ( \hat{\Pi}_I )_w ( \hat{A} )_{w'}\Tr_{A} ((\hat{\pi}_{A}^l \otimes \mathbb{I}) \rho_{AB}) \nonumber\\
	&& - ig  ( \hat{\Pi}_I )_w^* \Tr_{A}( (\hat{\pi}_{A}^l \otimes \mathbb{I}) \rho_{AB} ( \hat{A} \otimes \mathbb{I})) + ig ( \hat{\Pi}_I )_w \Tr_{A} ( (\hat{\pi}_{A}^l \otimes \mathbb{I}) (\hat{A} \otimes \mathbb{I}) \rho_{AB} )). 
	\end{eqnarray}
	\item Bob's expectation value:
	\begin{eqnarray}
	\label{B5}
	\langle \hat{B} \rangle_f^{Re} &=& \Tr (\hat{B}\rho_B^{N}) \nonumber\\
	&\approx& \frac{1}{\Tr ((\hat{\pi}_{A}^l \otimes \mathbb{I}) \rho_{AB})} \bigg[ \Tr ((\hat{\pi}_{A}^l \otimes \mathbb{I}) (\mathbb{I} \otimes \hat{B}) \rho_{AB}) + ig  ( \hat{\Pi}_I )_w^* ( \hat{A} )_{w'}^*\Tr ((\hat{\pi}_{A}^l \otimes \mathbb{I}) (\mathbb{I} \otimes \hat{B}) \rho_{AB}) \nonumber\\
	&& - ig ( \hat{\Pi}_I )_w ( \hat{A} )_{w'}\Tr ((\hat{\pi}_{A}^l \otimes \mathbb{I}) (\mathbb{I} \otimes \hat{B}) \rho_{AB}) - ig ( \hat{\Pi}_I )_w^* \Tr ( (\hat{\pi}_{A}^l \otimes \mathbb{I}) (\mathbb{I} \otimes \hat{B}) \rho_{AB} ( \hat{A} \otimes \mathbb{I})) \nonumber\\
	&& + ig ( \hat{\Pi}_I )_w \Tr ( (\hat{\pi}_{A}^l \otimes \mathbb{I}) (\mathbb{I} \otimes \hat{B}) (\hat{A} \otimes \mathbb{I}) \rho_{AB} ) \bigg] \nonumber\\
	&=& \frac{1}{\Tr ((\hat{\pi}_{A}^l \otimes \mathbb{I}) \rho_{AB})} \bigg[  \Tr ((\hat{\pi}_{A}^l \otimes \mathbb{I}) (\mathbb{I} \otimes \hat{B}) \rho_{AB}) - ig \Re ( \hat{\Pi}_I )_w (2 i \Im ( \hat{A} )_{w'}\Tr ((\hat{\pi}_{A}^l \otimes \mathbb{I}) (\mathbb{I} \otimes \hat{B}) \rho_{AB}) \nonumber\\
	&& - \Tr ((\hat{\pi}_{A}^l \otimes \mathbb{I}) (\mathbb{I} \otimes \hat{B}) [(\hat{A} \otimes \mathbb{I}), \rho_{AB}] )) + g \Im ( \hat{\Pi}_I )_w ( 2 \Re ( \hat{A} )_{w'} \Tr ((\hat{\pi}_{A}^l \otimes \mathbb{I}) (\mathbb{I} \otimes \hat{B}) \rho_{AB}) + \nonumber\\
	&& \Tr ((\hat{\pi}_{A}^l \otimes \mathbb{I}) (\mathbb{I} \otimes \hat{B}) \{ (\hat{A} \otimes \mathbb{I}), \rho_{AB} \} ))\bigg], 
	\end{eqnarray}
	where $ [x,y] $ and $\{x,y\} $ denote the commutator and the anticommutator respectively.
\end{enumerate}

\section{Algebraic details -- Proof}
\label{appC}
\begin{enumerate}
	\item Now we proceed to prove that denominators of the real and imaginary parts of the weak value do not go to zero for generic choices of $ \hat{A} $ and $ \hat{B} $. 
	Considering the extreme case, when all of the $ \rho_{ij} = 0 $ for a $ 2 \otimes 2 $ resource state, $ \hat{B} = \begin{bmatrix}
		b_{11} & b_{12} \\
		b_{12}^* & b_{22}
	\end{bmatrix} $, $ \hat{A} = \begin{bmatrix}
		a_{11} & a_{12} \\
		a_{12}^* & a_{22}
	\end{bmatrix} $, $ \hat{\pi}^l_A = \ket{\psi} \bra{\psi} $ (where $ \ket{\psi} = p_1 \ket{0} + p_2 \ket{1} $), $ p_1 \equiv p_{1R} + i p_{1I} $, $ p_2 \equiv p_{2R} + i p_{2I} $, and $ a_{12} \equiv a_{12R} + i a_{12I} $, denominator of real part of the weak value is given by $ b_{22} g (a_{12} p_2 p_1 + (2 p_2 (a_{12I} p_{1I} p_{2I} + a_{12R} p_{1R} p_{2I} - a_{12R} p_{1I} p_{2R} + a_{12I} p_{1R} p_{2R}) - i p_1 a_{12}^*) i p_2^* $ and denominator of the imaginary part of the weak value is given by $ a_{22} b_{22} - a_{22} b_{22}^2 - a_{22} b_{12} b_{12}^* $. Clearly, both of these are non-zero for any choice of observables $ \hat{A} $, $ \hat{B} $, and $ \hat{p}_1 $.

	\item For a $ 2 \cross 2 $ state $ \rho_{AB} $, we take $ \hat{B} $ to be a spin-1/2 observable $ \hat{n}.\hat{\sigma} $. Since $ \hat{B}^2 = 1 $ and $ \Tr_A( (\hat{A} \otimes \openone) \rho_{AB} ) = \Tr( \hat{A} \rho_{A}^{in} ) $, when we equate denominator of the imaginary part (Eq. \ref{im}) to zero, we find the condition $ \Tr( \hat{B} \rho_B^{in} ) = 1 $. This is satisfied when Bob's reduced state $ \rho_{B}^{in} $ -- when it is pure -- is an eigenstate of $ \hat{B} $. So to ensure $ \Tr( \hat{B} \rho_B^{in} ) \neq 1 $, set of observables $ \hat{B} $ must exclude the observable whose ``up'' eigenstate is $ \rho_{B}^{in} $. When $ \rho_{B}^{in} $ is a mixed state, the observable set $ \hat{B} $ must be such that it has the elements of a complete observable basis set $ \{\hat{n_1}.\hat{\sigma}, \hat{n_2}.\hat{\sigma}, \hat{n_3}.\hat{\sigma}\} $ such that $ \hat{n_1} \perp \hat{n_2} \perp \hat{n_3}  $.

	\item Consider spin-1/2 observables $ B_\alpha = (1/\sqrt{n_{1\alpha}^2 + n_{2\alpha}^2 + n_{3\alpha}^2)} \vec{n_\alpha}.\vec{\sigma} $, $ B_\beta = (1/\sqrt{n_{1\beta}^2 + n_{2\beta}^2 + n_{3\beta}^2)} \vec{n_\beta}.\vec{\sigma} $, and $ B_\gamma = (1/\sqrt{n_{1\gamma}^2 + n_{2\gamma}^2 + n_{3\gamma}^2)} \vec{n_\gamma}.\vec{\sigma} $. Solving the set of equations $ \Tr(B_\alpha \rho) = \Tr(B_\alpha \rho'); \Tr(B_\beta \rho) = \Tr(B_\beta \rho'); \Tr(B_\gamma \rho) = \Tr(B_\gamma \rho') $ (where $ \rho $ and $ \rho' $ are general $ 2 \cross 2 $ density matrices) leads to $ \rho = \rho' $. This is easily extended to higher dimensions by considering higher dimensional versions of Pauli matrices~\cite{bertlmann2008bloch,stephany1979higher}.    	

	\item Substituting $ \rho_{AB} = \rho_{A} \otimes \rho_{B} $ in Eq.~(A2), which corresponds to the first set of experimental runs (imaginary part), implies:
	\begin{eqnarray}
	\label{C1}
	\rho_{B}^N = \frac{\rho_{B}^{un}}{\Tr(\rho_B^{un})} 
	&\approx& \frac{\rho_B - 2 g \Im ( \hat{\Pi}_I )_w \Tr_{A} ((\hat{A} \otimes \mathbb{I}) \rho_{A} \otimes \rho_{B}) }{ \bigg(\Tr(\rho_B) - 2 g \Im ( \hat{\Pi}_I )_w \Tr ((\hat{A} \otimes \mathbb{I}) \rho_{A} \otimes \rho_{B}) \bigg)} \nonumber\\
	&=& \frac{(1 - 2 g \Im ( \hat{\Pi}_I )_w \Tr_{A} (\hat{A} \rho_{A}) ) \rho_B}{(1 - 2 g \Im ( \hat{\Pi}_I )_w \Tr_A (\hat{A} \rho_{A} ) ) \Tr_B \rho_{B}} \nonumber\\
	&=& \rho_B. 
	\end{eqnarray}	
	\item For real part of the weak value, we consider 
	\begin{gather}
		\Re ( \hat{\Pi}_I )_w = \nonumber\\
		\langle \hat{B} \rangle^{Re}_f \cross \Tr((\hat{\pi}_A^l \otimes \openone) \rho_{AB}) - \Tr((\hat{\pi}_A^l \otimes \hat{B}) \rho_{AB}) - g \Im( \hat{\Pi}_I )_w \big(2 \Re ( \hat{A} )_{w'} \Tr((\hat{\pi}_A^l \otimes \hat{B}) \rho_{AB}) \nonumber\\
		\label{reprf1}
		+ \Tr((\hat{\pi}_A^l \otimes \hat{B}) \{(\hat{A} \otimes \openone),\rho_{AB}\} ) \big) = 0.
	\end{gather}
	This leads to 
	\begin{gather}
		\langle \hat{B} \rangle^{Re}_f
		= \frac{\Tr((\hat{\pi}_A^l \otimes \hat{B}) \rho_{AB})}{\Tr((\hat{\pi}_A^l \otimes \openone) \rho_{AB})}
		+ g \Im ( \hat{\Pi}_I )_w \big[ \frac{\Tr((\hat{\pi}_A^l \hat{A} \otimes \openone)\rho_{AB})}{\Tr((\hat{\pi}_A^l \otimes \openone) \rho_{AB})} + \frac{\Tr((\hat{A} \hat{\pi}_A^l \otimes \openone)\rho_{AB}^\dagger)}{\Tr((\hat{\pi}_A^l \otimes \openone) \rho_{AB}^\dagger)} \big] \frac{\Tr((\hat{\pi}_A^l \otimes \hat{B}) \rho_{AB})}{\Tr((\hat{\pi}_A^l \otimes \openone) \rho_{AB})} \nonumber\\
        \label{reprf2}			
		+ g \Im ( \hat{\Pi}_I )_w \big[\frac{\Tr((\hat{\pi}^l_A \hat{A} \otimes \hat{B}) \rho_{AB}) + \Tr((\hat{A} \hat{\pi}^l_A \otimes \hat{B}) \rho_{AB})}{\Tr((\hat{\pi}_A^l \otimes \openone) \rho_{AB})} \big],
	\end{gather}	
	For the $ l^{th} $ projector (total $p$ projectors on the Hilbert space of same dimensions), 
	$ E_l \equiv \Tr((\hat{\pi}_A^l \otimes \hat{B}) \rho_{AB}) $, 
	$ N_l \equiv \Tr((\hat{\pi}_A^l \otimes \openone) \rho_{AB}) $, 
	and $ E \equiv \Tr(\hat{B} \rho_B^{in}) = \Tr((\openone \otimes \hat{B}) \rho_{AB}) $. We have $ \sum_{l=1}^p E_l = E $. Substituting these entities in Equation.~\ref{reprf2} and replacing terms in the brackets corresponding to the projectors with $ T1_l $ and $ T2_l $ respectively, we find:
	\begin{eqnarray}
	\langle \hat{B} \rangle^{Re}_f &=& E_l/N_l + g \Im ( \hat{\Pi}_I )_w [T1_l] E_l/N_l + g \Im ( \hat{\Pi}_I )_w [T2_l] \nonumber\\
	\label{reprf3}
	&\implies& \sum_{l=1}^p N_l \langle \hat{B} \rangle^{Re}_f = \sum_{l=1}^p E_l + g \Im ( \hat{\Pi}_I )_w \sum_{l=1}^p [T1_l] E_l + g \Im ( \hat{\Pi}_I )_w \sum_{l=1}^p [T2_l] N_l
	\end{eqnarray}  
 	Since $ \sum_{l=1}^p = \hat{\pi}^l_A = \openone $, $ \sum_{l=1}^p N_l = \Tr(\rho_{AB}) = 1 $. Therefore, 
 	\begin{equation}
 	\label{reprf4}
 	\langle \hat{B} \rangle^{Re}_f = E + g \Im\langle \hat{\Pi}_I \rangle_w \sum_{l=1}^p [T1_l] E_l + g \Im \langle \hat{\Pi}_I \rangle_w \sum_{l=1}^p [T2_l] N_l
 	\end{equation} 
	\item Substituting $ \rho_{AB} = \rho_A \otimes \rho_B $ in Eq.~(B3), corresponding to the second set of experimental runs (real part), implies:
	\begin{gather}
	\rho_{B}^N = \frac{\rho_{B}^{un}}{\Tr(\rho_B^{un})} \approx \nonumber\\
	\frac{\Tr_{I} (\hat{\pi}_{I}^v \rho_{I}) ( ( \Tr_{A} (\hat{\pi}_{A}^l \rho_{A}) \rho_B  - ig  ( \hat{\Pi}_I )_w^* \Tr_{A}( \hat{\pi}_{A}^l \rho_{A} \hat{A} ) \rho_B + ig ( \hat{\Pi}_I )_w \Tr_{A} ( \hat{\pi}_{A}^l  \hat{A} \rho_{A} ) \rho_B ))}{\Tr_{I} (\hat{\pi}_{I}^v \rho_{I}) \Tr(\rho_B) ( ( \Tr_{A} (\hat{\pi}_{A}^l \rho_{A}) - ig  ( \hat{\Pi}_I )_w^* \Tr_{A}( \hat{\pi}_{A}^l \rho_{A} \hat{A} ) + ig ( \hat{\Pi}_I )_w \Tr_{A} ( \hat{\pi}_{A}^l \hat{A} \rho_{A} ) ))} \nonumber\\
	= \frac{ \Tr_{I} (\hat{\pi}_{I}^v \rho_{I}) ( ( \Tr_{A} (\hat{\pi}_{A}^l  \rho_{A}) - ig  ( \hat{\Pi}_I )_w^* \Tr_{A}( \hat{\pi}_{A}^l \rho_{A} \hat{A} ) + ig ( \hat{\Pi}_I )_w \Tr_{A} ( \hat{\pi}_{A}^l \hat{A} \rho_{A} ) )) \rho_B}{\Tr_{I} (\hat{\pi}_{I}^v \rho_{I}) ( ( \Tr_{A} (\hat{\pi}_{A}^l \rho_{A}) - ig  ( \hat{\Pi}_I )_w^* \Tr_{A}( \hat{\pi}_{A}^l \rho_{A} \hat{A} ) + ig ( \hat{\Pi}_I )_w \Tr_{A} ( \hat{\pi}_{A}^l \hat{A} \rho_{A} ) ))} \nonumber\\
	\label{C2}
	= \rho_B.
	\end{gather}   
	Like in case of the first set of experiments, here too, Bob's state contains no signature of the weak measurement performed by Alice if $ \rho_{AB} $ is a product state. 
\end{enumerate}

\section{Algebraic details -- Bell-diagonal state as resource}
\label{appD}
\begin{enumerate}
	\item Imaginary part:
	\begin{equation}
	\label{D1}
	\Im ( \hat{\Pi}_I )_w = \frac{b_{11} + b_{22} - 2 \langle \hat{B} \rangle_f^{Im} }{ \left( \splitfrac{g[(a_{11} + a_{22})[(b_{11}-1)b_{11} + (b_{22}-1)b_{22} + 2(a_{11}+a_{22})|b_{12}|^2 +}{(b_{11} + b_{22})(4 \Re(a_{12}) \Re(b_{12})c_1 + 4 \Im(a_{12}) \Im(b_{12}) c_2 + (a_{11} - a_{22}) (b_{11} - b_{22}) c_3)]]} \right)}.
	\end{equation}
	Here, $ \langle  \rangle^{Im} $ represents expectation value obtained in the set of experiments which correspond to obtaining the imaginary part of the weak value.	
	\item Real part:
	\begin{gather}
	\Re ( \hat{\Pi}_I )_w = \nonumber\\
	\label{D2}
	\dfrac{ \left( \splitdfrac{2 \langle \hat{B} \rangle^{Re}_f + b_{11} (c_3 - 4 a_{22} g \Im \langle \hat{\Pi}_I \rangle_w - 1) - b_{22} (c_3 + 4 a_{22} g \Im \langle \hat{\Pi}_I \rangle_w + 1)}{- 4 g \Im \langle \hat{\Pi}_I \rangle_w [\Re(a_{12}) \Re(b_{12}) c_1 + \Im(a_{12}) \Im(b_{12}) c_2 + c_3 a_{22} (b_{22} - b_{11})] }  \right) }{4g(\Im(a_{12}) \Re(b_{12}) c_1 + i \Re(a_{12}) (b_{12} - \Re(b_{12})) c_2)}.
	\end{gather}
	Here too, $ \langle \rangle^{Re} $ represents expectation value obtained in the set of experiments which correspond to obtaining the real part of the weak value. 
	\item Number of classical bits to be communicated:  \\
	The total state after weak interaction is $ \rho_{twk} = U \rho_{I} \otimes \rho_{AB} U^\dagger \approx \rho_{I} \otimes \rho_{AB} + ig [\ket{a_k}\bra{a_k}, \rho_{I}] \otimes [\hat{A} \otimes \mathbb{I}, \rho_{AB}] $. Similarly, $ \rho_{twk}^2 = U \rho_{I}^2 \otimes \rho_{AB}^2 U^\dagger \approx \rho_{I}^2 \otimes \rho_{AB}^2 + ig \ket{a_k}\bra{a_k} \rho_{I}^2 \otimes ( \hat{A} \otimes \mathbb{I} \rho_{AB}^2 ) - ig \rho_{I}^2 \ket{a_k}\bra{a_k} \otimes ( \rho_{AB}^2 \hat{A} \otimes \mathbb{I}) $. Here, we have chosen $ \hat{\Pi}^k_I = \ket{a_k}\bra{a_k} $. Substituting $ \hat{\pi}_{I}^v = \ket{b_0} \bra{b_0} = (1/d) \sum_{a,b = 0}^{d-1} \ket{a}\bra{b} $ and $ \rho_{I}^2 = \rho_{I} = \sum_{a,b = 0}^{d-1} \psi_a \psi_b^\ast \ket{a}\bra{b} $ into Equation~\ref{cbits}, where the first and second terms in the parenthesis correspond to post-selection probabilities for the first and second set of experimental runs respectively, we get:
	\begin{eqnarray}
	C &=& N \sum_{k = 0}^{d-1} \bigg[ \Tr ( \hat{\pi}_{I}^v \rho_{twk}^2 ) + \Tr ( ( \hat{\pi}_{I}^v \otimes \hat{\pi}_{A}^l \otimes \mathbb{I}) \rho_{twk}^2 ) \bigg] \nonumber\\
	&\approx& N \sum_{k = 0}^{d-1} \bigg[ \Tr ( \hat{\pi}_{I}^v \rho_I \otimes \rho_{AB}^2 + i g \hat{\pi}_{I}^v \ket{a_k}\bra{a_k} \rho_{I} \otimes ( \hat{A} \otimes \mathbb{I} \rho_{AB}^2 ) - i g \hat{\pi}_{I}^v \rho_{I} \ket{a_k}\bra{a_k} \otimes ( \rho_{AB}^2 \hat{A} \otimes \mathbb{I} ) ) \nonumber\\
	&& + \Tr ( \hat{\pi}_{I}^v \rho_I \otimes ( (\hat{\pi}^l_A \otimes \mathbb{I}) \rho_{AB}^2 ) + ig \hat{\pi}_{I}^v \ket{a_k}\bra{a_k} \rho_{I} \otimes (\hat{\pi}_{A}^l \otimes \mathbb{I}) ( \hat{A} \otimes \mathbb{I} \rho_{AB}^2) - ig \hat{\pi}_{I}^v \rho_{I} \ket{a_k}\bra{a_k} \otimes (\hat{\pi}_{A}^l \otimes \mathbb{I}) ( \rho_{AB}^2 \hat{A} \otimes \mathbb{I}) ) \bigg] \nonumber\\
	&=& N \sum_{k = 0}^{d-1} \bigg[ \Tr ( \hat{\pi}_{I}^v \rho_I ) \Tr ( \rho_{AB}^2 ) + i g \Tr ( \hat{\pi}_{I}^v \ket{a_k}\bra{a_k} \rho_{I} ) \Tr ( \hat{A} \otimes \mathbb{I} \rho_{AB}^2 ) - i g \Tr ( \hat{\pi}_{I}^v \rho_{I} \ket{a_k}\bra{a_k} ) \Tr ( \rho_{AB}^2 \hat{A} \otimes \mathbb{I} )  \nonumber\\
	&& + \Tr ( \hat{\pi}_{I}^v \rho_I ) \Tr ( (\hat{\pi}^l_A \otimes \mathbb{I}) \rho_{AB}^2 ) + ig \Tr ( \hat{\pi}_{I}^v \ket{a_k}\bra{a_k} \rho_{I} ) \Tr (\hat{\pi}_{A}^l \hat{A} \otimes \mathbb{I} \rho_{AB}^2) \nonumber\\
	&&  - ig \Tr ( \hat{\pi}_{I}^v \rho_{I} \ket{a_k}\bra{a_k} ) \Tr ( (\hat{\pi}_{A}^l \otimes \mathbb{I}) ( \rho_{AB}^2 \hat{A} \otimes \mathbb{I}) ) \bigg] \nonumber\\
	&=& N \bigg[ \sum_{k = 0}^{d-1} ( \Tr ( \hat{\pi}_{I}^v \rho_I ) \Tr ( \rho_{AB}^2 ) ) + i g \Tr ( \hat{\pi}_{I}^v \sum_{k = 0}^{d-1} \ket{a_k}\bra{a_k} \rho_{I} ) \Tr ( \hat{A} \otimes \mathbb{I} \rho_{AB}^2 ) - i g \Tr ( \hat{\pi}_{I}^v \rho_{I} \sum_{k = 0}^{d-1} \ket{a_k}\bra{a_k} ) \Tr ( \rho_{AB}^2 \hat{A} \otimes \mathbb{I} )  \nonumber\\
	&& + \sum_{k = 0}^{d-1} \Tr ( \hat{\pi}_{I}^v \rho_I ) \Tr ( (\hat{\pi}^l_A \otimes \mathbb{I}) \rho_{AB}^2 ) + ig \Tr ( \hat{\pi}_{I}^v \sum_{k = 0}^{d-1} \ket{a_k}\bra{a_k} \rho_{I} ) \Tr (\hat{\pi}_{A}^l \hat{A} \otimes \mathbb{I} \rho_{AB}^2) \nonumber\\
	\label{D3}
	&&  - ig \Tr ( \hat{\pi}_{I}^v \rho_{I} \sum_{k = 0}^{d-1} \ket{a_k}\bra{a_k} ) \Tr ( (\hat{\pi}_{A}^l \otimes \mathbb{I}) ( \rho_{AB}^2 \hat{A} \otimes \mathbb{I}) ) \bigg];
	\end{eqnarray}
	since $\sum_{k = 0}^{d-1} \ket{a_k}\bra{a_k} = \mathbb{I}$ and $\sum_{k = 0}^{d-1} 1 = d $, we have,
	\begin{eqnarray}
	\label{D4}
	C &\approx& N \bigg[ d \Tr ( \hat{\pi}_{I}^v \rho_I ) \Tr ( \rho_{AB}^2 ) + i g \Tr ( \hat{\pi}_{I}^v \rho_{I} ) \Tr ( \hat{A} \otimes \mathbb{I} \rho_{AB}^2 ) - i g \Tr ( \hat{\pi}_{I}^v \rho_{I} ) \Tr ( \rho_{AB}^2 \hat{A} \otimes \mathbb{I} )  \nonumber\\
	&& + d \Tr ( \hat{\pi}_{I}^v \rho_I ) \Tr ( (\hat{\pi}^l_A \otimes\mathbb{I}) \rho_{AB}^2 ) + ig \Tr ( \hat{\pi}_{I}^v \rho_{I} ) \Tr (\hat{\pi}_{A}^l \hat{A} \otimes \mathbb{I} \rho_{AB}^2) \nonumber\\
	&&  - ig \Tr ( \hat{\pi}_{I}^v \rho_{I} ) \Tr ( (\hat{\pi}_{A}^l \otimes \mathbb{I}) ( \rho_{AB}^2 \hat{A} \otimes \mathbb{I}) ) \bigg]. 
	\end{eqnarray}
	Substituting $ \hat{A} = \begin{bmatrix}
		a_{11} & a_{12} \\
		a_{12}* & a_{22} 
	\end{bmatrix} $, and $ \hat{\pi}_A^l = \ket{\sigma_{zA} = -1}\bra{\sigma_{zA} = -1} $, we get
	\begin{eqnarray}
	\label{D5}
	C &\approx& N [d \Tr ( \hat{\pi}_{I}^v \rho_I ) \Tr ( \rho_{AB}^2 ) + (d/2) \Tr ( \hat{\pi}_{I}^v \rho_I ) \Tr ( \rho_{AB}^2 )] \nonumber\\
	&=& (3/2) N d \Tr ( \hat{\pi}_{I}^v \rho_I ) \Tr ( \rho_{AB}^2 ).
	\end{eqnarray}
	The post-selection probabilities corresponding to the first and second runs of the protocol are represented by the first and second terms in the parenthesis respectively. As expected from the simultaneous success probability requirement, in the second set of experimental runs, post-selection succeeds exactly half the number of times it does in the first set. Considering the system state to be transferred as $ \rho_I = \ket{\psi}\bra{\psi} = \sum_{k=0}^{d-1} a_k a_k^* \ket{a_k} \bra{a_k} $, we have $ C \propto \Tr(\hat{\pi}^v_I \rho_I) = \Tr(\ket{b_0} \bra{b_0} \sum_{k=0}^{d-1} a_k^* a_l \ket{a_l} \bra{a_k} ) = \Tr((1/d) \sum_{a,b = 0}^{d-1} \ket{a} \bra{b} \sum_{k,l=0}^{d-1} a_k^* a_l \ket{a_l} \bra{a_k} ) = (1/d) \sum_{a,b=0}^{d-1} \sum_{k=0}^{d-1} a_k^* a_l \braket{b}{a_l} \braket{a_k}{a} = (1/d) \sum_{k=0}^{d-1} \sum_{a,b=0}^{d-1} a_k^* a_l \delta_{b a_l} \delta{a_k a} = \sum_{a,b = 0}^{d - 1} a_b^* a_a = 0 $ (because $ \braket{b}{a_k} = \delta_{b a_k} $ and $ \braket{a_k}{a} = \delta_{a a_k} $). The solution space for $ \sum_{k = 0}^{d - 1} a_k = 0 $ is negligible compared to rest of the possibilities. Therefore, the success probability is unlikely to go to zero for any state of interest that is to be transferred. 
	
\end{enumerate}

\newpage
\bibliographystyle{apsrev4-1}
\bibliography{RSD}

\begin{thebibliography}{82}%
\makeatletter
\providecommand \@ifxundefined [1]{%
 \@ifx{#1\undefined}
}%
\providecommand \@ifnum [1]{%
 \ifnum #1\expandafter \@firstoftwo
 \else \expandafter \@secondoftwo
 \fi
}%
\providecommand \@ifx [1]{%
 \ifx #1\expandafter \@firstoftwo
 \else \expandafter \@secondoftwo
 \fi
}%
\providecommand \natexlab [1]{#1}%
\providecommand \enquote  [1]{``#1''}%
\providecommand \bibnamefont  [1]{#1}%
\providecommand \bibfnamefont [1]{#1}%
\providecommand \citenamefont [1]{#1}%
\providecommand \href@noop [0]{\@secondoftwo}%
\providecommand \href [0]{\begingroup \@sanitize@url \@href}%
\providecommand \@href[1]{\@@startlink{#1}\@@href}%
\providecommand \@@href[1]{\endgroup#1\@@endlink}%
\providecommand \@sanitize@url [0]{\catcode `\\12\catcode `\$12\catcode
  `\&12\catcode `\#12\catcode `\^12\catcode `\_12\catcode `\%12\relax}%
\providecommand \@@startlink[1]{}%
\providecommand \@@endlink[0]{}%
\providecommand \url  [0]{\begingroup\@sanitize@url \@url }%
\providecommand \@url [1]{\endgroup\@href {#1}{\urlprefix }}%
\providecommand \urlprefix  [0]{URL }%
\providecommand \Eprint [0]{\href }%
\providecommand \doibase [0]{http://dx.doi.org/}%
\providecommand \selectlanguage [0]{\@gobble}%
\providecommand \bibinfo  [0]{\@secondoftwo}%
\providecommand \bibfield  [0]{\@secondoftwo}%
\providecommand \translation [1]{[#1]}%
\providecommand \BibitemOpen [0]{}%
\providecommand \bibitemStop [0]{}%
\providecommand \bibitemNoStop [0]{.\EOS\space}%
\providecommand \EOS [0]{\spacefactor3000\relax}%
\providecommand \BibitemShut  [1]{\csname bibitem#1\endcsname}%
\let\auto@bib@innerbib\@empty
\bibitem [{\citenamefont {Bennett}\ and\ \citenamefont
  {Brassard}(2014)}]{bennett2014quantum}%
  \BibitemOpen
  \bibfield  {author} {\bibinfo {author} {\bibfnamefont {C.~H.}\ \bibnamefont
  {Bennett}}\ and\ \bibinfo {author} {\bibfnamefont {G.}~\bibnamefont
  {Brassard}},\ }\href@noop {} {\bibfield  {journal} {\bibinfo  {journal}
  {Theoretical Computer Science}\ }\textbf {\bibinfo {volume} {560}},\ \bibinfo
  {pages} {7} (\bibinfo {year} {2014})}\BibitemShut {NoStop}%
\bibitem [{\citenamefont {Einstein}\ \emph {et~al.}(1935)\citenamefont
  {Einstein}, \citenamefont {Podolsky},\ and\ \citenamefont
  {Rosen}}]{einstein1935can}%
  \BibitemOpen
  \bibfield  {author} {\bibinfo {author} {\bibfnamefont {A.}~\bibnamefont
  {Einstein}}, \bibinfo {author} {\bibfnamefont {B.}~\bibnamefont {Podolsky}},
  \ and\ \bibinfo {author} {\bibfnamefont {N.}~\bibnamefont {Rosen}},\
  }\href@noop {} {\bibfield  {journal} {\bibinfo  {journal} {Physical review}\
  }\textbf {\bibinfo {volume} {47}},\ \bibinfo {pages} {777} (\bibinfo {year}
  {1935})}\BibitemShut {NoStop}%
\bibitem [{\citenamefont {Schr{\"o}dinger}(1935)}]{schrodinger1935discussion}%
  \BibitemOpen
  \bibfield  {author} {\bibinfo {author} {\bibfnamefont {E.}~\bibnamefont
  {Schr{\"o}dinger}},\ }in\ \href@noop {} {\emph {\bibinfo {booktitle}
  {Mathematical Proceedings of the Cambridge Philosophical Society}}},\
  Vol.~\bibinfo {volume} {31}\ (\bibinfo {organization} {Cambridge Univ
  Press},\ \bibinfo {year} {1935})\ pp.\ \bibinfo {pages}
  {555--563}\BibitemShut {NoStop}%
\bibitem [{\citenamefont {Ekert}(1991)}]{ekert1991quantum}%
  \BibitemOpen
  \bibfield  {author} {\bibinfo {author} {\bibfnamefont {A.~K.}\ \bibnamefont
  {Ekert}},\ }\href@noop {} {\bibfield  {journal} {\bibinfo  {journal}
  {Physical review letters}\ }\textbf {\bibinfo {volume} {67}},\ \bibinfo
  {pages} {661} (\bibinfo {year} {1991})}\BibitemShut {NoStop}%
\bibitem [{\citenamefont {Bennett}\ \emph {et~al.}(1993)\citenamefont
  {Bennett}, \citenamefont {Brassard}, \citenamefont {Cr{\'e}peau},
  \citenamefont {Jozsa}, \citenamefont {Peres},\ and\ \citenamefont
  {Wootters}}]{bennett1993teleporting}%
  \BibitemOpen
  \bibfield  {author} {\bibinfo {author} {\bibfnamefont {C.~H.}\ \bibnamefont
  {Bennett}}, \bibinfo {author} {\bibfnamefont {G.}~\bibnamefont {Brassard}},
  \bibinfo {author} {\bibfnamefont {C.}~\bibnamefont {Cr{\'e}peau}}, \bibinfo
  {author} {\bibfnamefont {R.}~\bibnamefont {Jozsa}}, \bibinfo {author}
  {\bibfnamefont {A.}~\bibnamefont {Peres}}, \ and\ \bibinfo {author}
  {\bibfnamefont {W.~K.}\ \bibnamefont {Wootters}},\ }\href@noop {} {\bibfield
  {journal} {\bibinfo  {journal} {Physical review letters}\ }\textbf {\bibinfo
  {volume} {70}},\ \bibinfo {pages} {1895} (\bibinfo {year}
  {1993})}\BibitemShut {NoStop}%
\bibitem [{\citenamefont {Pati}(2000)}]{pati2000minimum}%
  \BibitemOpen
  \bibfield  {author} {\bibinfo {author} {\bibfnamefont {A.~K.}\ \bibnamefont
  {Pati}},\ }\href@noop {} {\bibfield  {journal} {\bibinfo  {journal} {Physical
  Review A}\ }\textbf {\bibinfo {volume} {63}},\ \bibinfo {pages} {014302}
  (\bibinfo {year} {2000})}\BibitemShut {NoStop}%
\bibitem [{\citenamefont {Bennett}\ \emph {et~al.}(2001)\citenamefont
  {Bennett}, \citenamefont {DiVincenzo}, \citenamefont {Shor}, \citenamefont
  {Smolin}, \citenamefont {Terhal},\ and\ \citenamefont
  {Wootters}}]{bennett2001remote}%
  \BibitemOpen
  \bibfield  {author} {\bibinfo {author} {\bibfnamefont {C.~H.}\ \bibnamefont
  {Bennett}}, \bibinfo {author} {\bibfnamefont {D.~P.}\ \bibnamefont
  {DiVincenzo}}, \bibinfo {author} {\bibfnamefont {P.~W.}\ \bibnamefont
  {Shor}}, \bibinfo {author} {\bibfnamefont {J.~A.}\ \bibnamefont {Smolin}},
  \bibinfo {author} {\bibfnamefont {B.~M.}\ \bibnamefont {Terhal}}, \ and\
  \bibinfo {author} {\bibfnamefont {W.~K.}\ \bibnamefont {Wootters}},\
  }\href@noop {} {\bibfield  {journal} {\bibinfo  {journal} {Physical Review
  Letters}\ }\textbf {\bibinfo {volume} {87}},\ \bibinfo {pages} {077902}
  (\bibinfo {year} {2001})}\BibitemShut {NoStop}%
\bibitem [{\citenamefont {Scarani}\ \emph {et~al.}(2009)\citenamefont
  {Scarani}, \citenamefont {Bechmann-Pasquinucci}, \citenamefont {Cerf},
  \citenamefont {Dušek}, \citenamefont {Lütkenhaus},\ and\ \citenamefont
  {Peev}}]{scarani2009the}%
  \BibitemOpen
  \bibfield  {author} {\bibinfo {author} {\bibfnamefont {V.}~\bibnamefont
  {Scarani}}, \bibinfo {author} {\bibfnamefont {H.}~\bibnamefont
  {Bechmann-Pasquinucci}}, \bibinfo {author} {\bibfnamefont {N.~J.}\
  \bibnamefont {Cerf}}, \bibinfo {author} {\bibfnamefont {M.}~\bibnamefont
  {Dušek}}, \bibinfo {author} {\bibfnamefont {N.}~\bibnamefont {Lütkenhaus}},
  \ and\ \bibinfo {author} {\bibfnamefont {M.}~\bibnamefont {Peev}},\
  }\href@noop {} {\bibfield  {journal} {\bibinfo  {journal} {Reviews of Modern
  Physics}\ }\textbf {\bibinfo {volume} {81}},\ \bibinfo {pages} {1302}
  (\bibinfo {year} {2009})}\BibitemShut {NoStop}%
\bibitem [{\citenamefont {Vaidman}(1994)}]{vaidman1994teleportation}%
  \BibitemOpen
  \bibfield  {author} {\bibinfo {author} {\bibfnamefont {L.}~\bibnamefont
  {Vaidman}},\ }\href@noop {} {\bibfield  {journal} {\bibinfo  {journal}
  {Physical Review A}\ }\textbf {\bibinfo {volume} {49}},\ \bibinfo {pages}
  {1473} (\bibinfo {year} {1994})}\BibitemShut {NoStop}%
\bibitem [{\citenamefont {Braunstein}\ and\ \citenamefont
  {Kimble}(1998)}]{braunstein1998teleportation}%
  \BibitemOpen
  \bibfield  {author} {\bibinfo {author} {\bibfnamefont {S.~L.}\ \bibnamefont
  {Braunstein}}\ and\ \bibinfo {author} {\bibfnamefont {H.~J.}\ \bibnamefont
  {Kimble}},\ }\href@noop {} {\bibfield  {journal} {\bibinfo  {journal}
  {Physical Review Letters}\ }\textbf {\bibinfo {volume} {80}},\ \bibinfo
  {pages} {869} (\bibinfo {year} {1998})}\BibitemShut {NoStop}%
\bibitem [{\citenamefont {Koniorczyk}\ \emph {et~al.}(2001)\citenamefont
  {Koniorczyk}, \citenamefont {Bu{\v{z}}ek},\ and\ \citenamefont
  {Janszky}}]{koniorczyk2001wigner}%
  \BibitemOpen
  \bibfield  {author} {\bibinfo {author} {\bibfnamefont {M.}~\bibnamefont
  {Koniorczyk}}, \bibinfo {author} {\bibfnamefont {V.}~\bibnamefont
  {Bu{\v{z}}ek}}, \ and\ \bibinfo {author} {\bibfnamefont {J.}~\bibnamefont
  {Janszky}},\ }\href@noop {} {\bibfield  {journal} {\bibinfo  {journal}
  {Physical Review A}\ }\textbf {\bibinfo {volume} {64}},\ \bibinfo {pages}
  {034301} (\bibinfo {year} {2001})}\BibitemShut {NoStop}%
\bibitem [{\citenamefont {Pirandola}\ \emph {et~al.}(2015)\citenamefont
  {Pirandola}, \citenamefont {Eisert}, \citenamefont {Weedbrook}, \citenamefont
  {Furusawa},\ and\ \citenamefont {Braunstein}}]{pirandola2015advances}%
  \BibitemOpen
  \bibfield  {author} {\bibinfo {author} {\bibfnamefont {S.}~\bibnamefont
  {Pirandola}}, \bibinfo {author} {\bibfnamefont {J.}~\bibnamefont {Eisert}},
  \bibinfo {author} {\bibfnamefont {C.}~\bibnamefont {Weedbrook}}, \bibinfo
  {author} {\bibfnamefont {A.}~\bibnamefont {Furusawa}}, \ and\ \bibinfo
  {author} {\bibfnamefont {S.}~\bibnamefont {Braunstein}},\ }\href@noop {}
  {\bibfield  {journal} {\bibinfo  {journal} {Nature Photonics}\ }\textbf
  {\bibinfo {volume} {9}},\ \bibinfo {pages} {641} (\bibinfo {year}
  {2015})}\BibitemShut {NoStop}%
\bibitem [{\citenamefont {Yonezawa}\ \emph {et~al.}(2004)\citenamefont
  {Yonezawa}, \citenamefont {Aoki},\ and\ \citenamefont
  {Furusawa}}]{yonezawa2004demonstration}%
  \BibitemOpen
  \bibfield  {author} {\bibinfo {author} {\bibfnamefont {H.}~\bibnamefont
  {Yonezawa}}, \bibinfo {author} {\bibfnamefont {T.}~\bibnamefont {Aoki}}, \
  and\ \bibinfo {author} {\bibfnamefont {A.}~\bibnamefont {Furusawa}},\
  }\href@noop {} {\bibfield  {journal} {\bibinfo  {journal} {Nature}\ }\textbf
  {\bibinfo {volume} {431}},\ \bibinfo {pages} {430} (\bibinfo {year}
  {2004})}\BibitemShut {NoStop}%
\bibitem [{\citenamefont {Andersen}\ and\ \citenamefont
  {Ralph}(2013)}]{andersen2013high}%
  \BibitemOpen
  \bibfield  {author} {\bibinfo {author} {\bibfnamefont {U.~L.}\ \bibnamefont
  {Andersen}}\ and\ \bibinfo {author} {\bibfnamefont {T.~C.}\ \bibnamefont
  {Ralph}},\ }\href@noop {} {\bibfield  {journal} {\bibinfo  {journal}
  {Physical review letters}\ }\textbf {\bibinfo {volume} {111}},\ \bibinfo
  {pages} {050504} (\bibinfo {year} {2013})}\BibitemShut {NoStop}%
\bibitem [{\citenamefont {Ulanov}\ \emph {et~al.}(2017)\citenamefont {Ulanov},
  \citenamefont {Sychev}, \citenamefont {Pushkina}, \citenamefont {Fedorov},\
  and\ \citenamefont {Lvovsky}}]{PhysRevLett.118.160501}%
  \BibitemOpen
  \bibfield  {author} {\bibinfo {author} {\bibfnamefont {A.~E.}\ \bibnamefont
  {Ulanov}}, \bibinfo {author} {\bibfnamefont {D.}~\bibnamefont {Sychev}},
  \bibinfo {author} {\bibfnamefont {A.~A.}\ \bibnamefont {Pushkina}}, \bibinfo
  {author} {\bibfnamefont {I.~A.}\ \bibnamefont {Fedorov}}, \ and\ \bibinfo
  {author} {\bibfnamefont {A.~I.}\ \bibnamefont {Lvovsky}},\ }\href {\doibase
  10.1103/PhysRevLett.118.160501} {\bibfield  {journal} {\bibinfo  {journal}
  {Phys. Rev. Lett.}\ }\textbf {\bibinfo {volume} {118}},\ \bibinfo {pages}
  {160501} (\bibinfo {year} {2017})}\BibitemShut {NoStop}%
\bibitem [{\citenamefont {Joo}\ \emph {et~al.}(2003)\citenamefont {Joo},
  \citenamefont {Park}, \citenamefont {Oh},\ and\ \citenamefont
  {Kim}}]{joo2003quantum}%
  \BibitemOpen
  \bibfield  {author} {\bibinfo {author} {\bibfnamefont {J.}~\bibnamefont
  {Joo}}, \bibinfo {author} {\bibfnamefont {Y.-J.}\ \bibnamefont {Park}},
  \bibinfo {author} {\bibfnamefont {S.}~\bibnamefont {Oh}}, \ and\ \bibinfo
  {author} {\bibfnamefont {J.}~\bibnamefont {Kim}},\ }\href@noop {} {\bibfield
  {journal} {\bibinfo  {journal} {New Journal of Physics}\ }\textbf {\bibinfo
  {volume} {5}},\ \bibinfo {pages} {136} (\bibinfo {year} {2003})}\BibitemShut
  {NoStop}%
\bibitem [{\citenamefont {Luo}\ \emph {et~al.}(2013)\citenamefont {Luo},
  \citenamefont {Li}, \citenamefont {Ma}, \citenamefont {Chen},\ and\
  \citenamefont {Yang}}]{luo2013faithful}%
  \BibitemOpen
  \bibfield  {author} {\bibinfo {author} {\bibfnamefont {M.-X.}\ \bibnamefont
  {Luo}}, \bibinfo {author} {\bibfnamefont {L.}~\bibnamefont {Li}}, \bibinfo
  {author} {\bibfnamefont {S.-Y.}\ \bibnamefont {Ma}}, \bibinfo {author}
  {\bibfnamefont {X.-B.}\ \bibnamefont {Chen}}, \ and\ \bibinfo {author}
  {\bibfnamefont {Y.-X.}\ \bibnamefont {Yang}},\ }\href@noop {} {\bibfield
  {journal} {\bibinfo  {journal} {International Journal of Theoretical
  Physics}\ }\textbf {\bibinfo {volume} {52}},\ \bibinfo {pages} {3032}
  (\bibinfo {year} {2013})}\BibitemShut {NoStop}%
\bibitem [{\citenamefont {Zhao}\ \emph {et~al.}(2011)\citenamefont {Zhao},
  \citenamefont {Li}, \citenamefont {Fei}, \citenamefont {Wang},\ and\
  \citenamefont {Li-Jost}}]{zhao2011faithful}%
  \BibitemOpen
  \bibfield  {author} {\bibinfo {author} {\bibfnamefont {M.-J.}\ \bibnamefont
  {Zhao}}, \bibinfo {author} {\bibfnamefont {Z.-G.}\ \bibnamefont {Li}},
  \bibinfo {author} {\bibfnamefont {S.-M.}\ \bibnamefont {Fei}}, \bibinfo
  {author} {\bibfnamefont {Z.-X.}\ \bibnamefont {Wang}}, \ and\ \bibinfo
  {author} {\bibfnamefont {X.}~\bibnamefont {Li-Jost}},\ }\href@noop {}
  {\bibfield  {journal} {\bibinfo  {journal} {Journal of Physics A:
  Mathematical and Theoretical}\ }\textbf {\bibinfo {volume} {44}},\ \bibinfo
  {pages} {215302} (\bibinfo {year} {2011})}\BibitemShut {NoStop}%
\bibitem [{\citenamefont {Albeverio}\ \emph {et~al.}(2002)\citenamefont
  {Albeverio}, \citenamefont {Fei},\ and\ \citenamefont
  {Yang}}]{albeverio2002optimal}%
  \BibitemOpen
  \bibfield  {author} {\bibinfo {author} {\bibfnamefont {S.}~\bibnamefont
  {Albeverio}}, \bibinfo {author} {\bibfnamefont {S.-M.}\ \bibnamefont {Fei}},
  \ and\ \bibinfo {author} {\bibfnamefont {W.-L.}\ \bibnamefont {Yang}},\
  }\href@noop {} {\bibfield  {journal} {\bibinfo  {journal} {Physical Review
  A}\ }\textbf {\bibinfo {volume} {66}},\ \bibinfo {pages} {012301} (\bibinfo
  {year} {2002})}\BibitemShut {NoStop}%
\bibitem [{\citenamefont {Cavalcanti}\ \emph {et~al.}(2017)\citenamefont
  {Cavalcanti}, \citenamefont {Skrzypczyk},\ and\ \citenamefont {\ifmmode
  \check{S}\else \v{S}\fi{}upi\ifmmode~\acute{c}\else
  \'{c}\fi{}}}]{PhysRevLett.119.110501}%
  \BibitemOpen
  \bibfield  {author} {\bibinfo {author} {\bibfnamefont {D.}~\bibnamefont
  {Cavalcanti}}, \bibinfo {author} {\bibfnamefont {P.}~\bibnamefont
  {Skrzypczyk}}, \ and\ \bibinfo {author} {\bibfnamefont {I.}~\bibnamefont
  {\ifmmode \check{S}\else \v{S}\fi{}upi\ifmmode~\acute{c}\else \'{c}\fi{}}},\
  }\href {\doibase 10.1103/PhysRevLett.119.110501} {\bibfield  {journal}
  {\bibinfo  {journal} {Phys. Rev. Lett.}\ }\textbf {\bibinfo {volume} {119}},\
  \bibinfo {pages} {110501} (\bibinfo {year} {2017})}\BibitemShut {NoStop}%
\bibitem [{\citenamefont {Agrawal}\ and\ \citenamefont
  {Pati}(2002)}]{agrawal2002probabilistic}%
  \BibitemOpen
  \bibfield  {author} {\bibinfo {author} {\bibfnamefont {P.}~\bibnamefont
  {Agrawal}}\ and\ \bibinfo {author} {\bibfnamefont {A.~K.}\ \bibnamefont
  {Pati}},\ }\href@noop {} {\bibfield  {journal} {\bibinfo  {journal} {Physics
  Letters A}\ }\textbf {\bibinfo {volume} {305}},\ \bibinfo {pages} {12}
  (\bibinfo {year} {2002})}\BibitemShut {NoStop}%
\bibitem [{\citenamefont {Biamonte}\ \emph {et~al.}(2017)\citenamefont
  {Biamonte}, \citenamefont {Wittek}, \citenamefont {Pancotti}, \citenamefont
  {Rebentrost}, \citenamefont {Wiebe},\ and\ \citenamefont
  {Lloyd}}]{biamonte2017quantum}%
  \BibitemOpen
  \bibfield  {author} {\bibinfo {author} {\bibfnamefont {J.}~\bibnamefont
  {Biamonte}}, \bibinfo {author} {\bibfnamefont {P.}~\bibnamefont {Wittek}},
  \bibinfo {author} {\bibfnamefont {N.}~\bibnamefont {Pancotti}}, \bibinfo
  {author} {\bibfnamefont {P.}~\bibnamefont {Rebentrost}}, \bibinfo {author}
  {\bibfnamefont {N.}~\bibnamefont {Wiebe}}, \ and\ \bibinfo {author}
  {\bibfnamefont {S.}~\bibnamefont {Lloyd}},\ }\href@noop {} {\bibfield
  {journal} {\bibinfo  {journal} {Nature}\ }\textbf {\bibinfo {volume} {549}},\
  \bibinfo {pages} {195} (\bibinfo {year} {2017})}\BibitemShut {NoStop}%
\bibitem [{\citenamefont {Diamanti}\ \emph {et~al.}(2016)\citenamefont
  {Diamanti}, \citenamefont {Lo}, \citenamefont {Qi},\ and\ \citenamefont
  {Yuanl}}]{diamanti2016practical}%
  \BibitemOpen
  \bibfield  {author} {\bibinfo {author} {\bibfnamefont {E.}~\bibnamefont
  {Diamanti}}, \bibinfo {author} {\bibfnamefont {H.-K.}\ \bibnamefont {Lo}},
  \bibinfo {author} {\bibfnamefont {B.}~\bibnamefont {Qi}}, \ and\ \bibinfo
  {author} {\bibfnamefont {Z.}~\bibnamefont {Yuanl}},\ }\href@noop {}
  {\bibfield  {journal} {\bibinfo  {journal} {npj Quantum Information}\
  }\textbf {\bibinfo {volume} {2}} (\bibinfo {year} {2016})}\BibitemShut
  {NoStop}%
\bibitem [{\citenamefont {Ma}\ \emph {et~al.}(2007)\citenamefont {Ma},
  \citenamefont {Fred~Fung},\ and\ \citenamefont {Lo}}]{ma2007quantum}%
  \BibitemOpen
  \bibfield  {author} {\bibinfo {author} {\bibfnamefont {X.}~\bibnamefont
  {Ma}}, \bibinfo {author} {\bibfnamefont {C.-H.}\ \bibnamefont {Fred~Fung}}, \
  and\ \bibinfo {author} {\bibfnamefont {H.-K.}\ \bibnamefont {Lo}},\
  }\href@noop {} {\bibfield  {journal} {\bibinfo  {journal} {Physics review A}\
  }\textbf {\bibinfo {volume} {76}} (\bibinfo {year} {2007})}\BibitemShut
  {NoStop}%
\bibitem [{\citenamefont {Lo}\ \emph {et~al.}(2014)\citenamefont {Lo},
  \citenamefont {Curty},\ and\ \citenamefont {Tamaki}}]{lo2014secure}%
  \BibitemOpen
  \bibfield  {author} {\bibinfo {author} {\bibfnamefont {H.-K.}\ \bibnamefont
  {Lo}}, \bibinfo {author} {\bibfnamefont {M.}~\bibnamefont {Curty}}, \ and\
  \bibinfo {author} {\bibfnamefont {K.}~\bibnamefont {Tamaki}},\ }\href@noop {}
  {\bibfield  {journal} {\bibinfo  {journal} {Nature Photonics}\ }\textbf
  {\bibinfo {volume} {8}},\ \bibinfo {pages} {595} (\bibinfo {year}
  {2014})}\BibitemShut {NoStop}%
\bibitem [{\citenamefont {Aharonov}\ \emph {et~al.}(1988)\citenamefont
  {Aharonov}, \citenamefont {Albert},\ and\ \citenamefont
  {Vaidman}}]{aharonov1988result}%
  \BibitemOpen
  \bibfield  {author} {\bibinfo {author} {\bibfnamefont {Y.}~\bibnamefont
  {Aharonov}}, \bibinfo {author} {\bibfnamefont {D.~Z.}\ \bibnamefont
  {Albert}}, \ and\ \bibinfo {author} {\bibfnamefont {L.}~\bibnamefont
  {Vaidman}},\ }\href@noop {} {\bibfield  {journal} {\bibinfo  {journal}
  {Physical review letters}\ }\textbf {\bibinfo {volume} {60}},\ \bibinfo
  {pages} {1351} (\bibinfo {year} {1988})}\BibitemShut {NoStop}%
\bibitem [{\citenamefont {Duck}\ \emph {et~al.}(1989)\citenamefont {Duck},
  \citenamefont {Stevenson},\ and\ \citenamefont
  {Sudarshan}}]{PhysRevD.40.2112}%
  \BibitemOpen
  \bibfield  {author} {\bibinfo {author} {\bibfnamefont {I.~M.}\ \bibnamefont
  {Duck}}, \bibinfo {author} {\bibfnamefont {P.~M.}\ \bibnamefont {Stevenson}},
  \ and\ \bibinfo {author} {\bibfnamefont {E.~C.~G.}\ \bibnamefont
  {Sudarshan}},\ }\href {\doibase 10.1103/PhysRevD.40.2112} {\bibfield
  {journal} {\bibinfo  {journal} {Phys. Rev. D}\ }\textbf {\bibinfo {volume}
  {40}},\ \bibinfo {pages} {2112} (\bibinfo {year} {1989})}\BibitemShut
  {NoStop}%
\bibitem [{\citenamefont {Kanjilal}\ \emph {et~al.}(2016)\citenamefont
  {Kanjilal}, \citenamefont {Muralidhara},\ and\ \citenamefont
  {Home}}]{kanjilal2016manifestation}%
  \BibitemOpen
  \bibfield  {author} {\bibinfo {author} {\bibfnamefont {S.}~\bibnamefont
  {Kanjilal}}, \bibinfo {author} {\bibfnamefont {G.}~\bibnamefont
  {Muralidhara}}, \ and\ \bibinfo {author} {\bibfnamefont {D.}~\bibnamefont
  {Home}},\ }\href@noop {} {\bibfield  {journal} {\bibinfo  {journal} {Physical
  Review A}\ }\textbf {\bibinfo {volume} {94}},\ \bibinfo {pages} {052110}
  (\bibinfo {year} {2016})}\BibitemShut {NoStop}%
\bibitem [{\citenamefont {Wu}\ and\ \citenamefont
  {M{\o}lmer}(2009)}]{wu2009weak}%
  \BibitemOpen
  \bibfield  {author} {\bibinfo {author} {\bibfnamefont {S.}~\bibnamefont
  {Wu}}\ and\ \bibinfo {author} {\bibfnamefont {K.}~\bibnamefont {M{\o}lmer}},\
  }\href@noop {} {\bibfield  {journal} {\bibinfo  {journal} {Physics Letters
  A}\ }\textbf {\bibinfo {volume} {374}},\ \bibinfo {pages} {34} (\bibinfo
  {year} {2009})}\BibitemShut {NoStop}%
\bibitem [{\citenamefont {Brun}\ \emph {et~al.}(2008)\citenamefont {Brun},
  \citenamefont {Di{\'o}si},\ and\ \citenamefont {Strunz}}]{brun2008test}%
  \BibitemOpen
  \bibfield  {author} {\bibinfo {author} {\bibfnamefont {T.~A.}\ \bibnamefont
  {Brun}}, \bibinfo {author} {\bibfnamefont {L.}~\bibnamefont {Di{\'o}si}}, \
  and\ \bibinfo {author} {\bibfnamefont {W.~T.}\ \bibnamefont {Strunz}},\
  }\href@noop {} {\bibfield  {journal} {\bibinfo  {journal} {Physical Review
  A}\ }\textbf {\bibinfo {volume} {77}},\ \bibinfo {pages} {032101} (\bibinfo
  {year} {2008})}\BibitemShut {NoStop}%
\bibitem [{\citenamefont {Lundeen}\ and\ \citenamefont
  {Resch}(2005)}]{lundeen2005practical}%
  \BibitemOpen
  \bibfield  {author} {\bibinfo {author} {\bibfnamefont {J.}~\bibnamefont
  {Lundeen}}\ and\ \bibinfo {author} {\bibfnamefont {K.}~\bibnamefont
  {Resch}},\ }\href@noop {} {\bibfield  {journal} {\bibinfo  {journal} {Physics
  Letters A}\ }\textbf {\bibinfo {volume} {334}},\ \bibinfo {pages} {337}
  (\bibinfo {year} {2005})}\BibitemShut {NoStop}%
\bibitem [{\citenamefont {Menzies}\ and\ \citenamefont
  {Korolkova}(2008)}]{menzies2008weak}%
  \BibitemOpen
  \bibfield  {author} {\bibinfo {author} {\bibfnamefont {D.}~\bibnamefont
  {Menzies}}\ and\ \bibinfo {author} {\bibfnamefont {N.}~\bibnamefont
  {Korolkova}},\ }\href@noop {} {\bibfield  {journal} {\bibinfo  {journal}
  {Physical Review A}\ }\textbf {\bibinfo {volume} {77}},\ \bibinfo {pages}
  {062105} (\bibinfo {year} {2008})}\BibitemShut {NoStop}%
\bibitem [{\citenamefont {Ho}\ \emph {et~al.}(2016)\citenamefont {Ho},
  \citenamefont {Boston}, \citenamefont {Palsson},\ and\ \citenamefont
  {Pryde}}]{ho2016experimental}%
  \BibitemOpen
  \bibfield  {author} {\bibinfo {author} {\bibfnamefont {J.}~\bibnamefont
  {Ho}}, \bibinfo {author} {\bibfnamefont {A.}~\bibnamefont {Boston}}, \bibinfo
  {author} {\bibfnamefont {M.}~\bibnamefont {Palsson}}, \ and\ \bibinfo
  {author} {\bibfnamefont {G.}~\bibnamefont {Pryde}},\ }\href@noop {}
  {\bibfield  {journal} {\bibinfo  {journal} {New Journal of Physics}\ }\textbf
  {\bibinfo {volume} {18}},\ \bibinfo {pages} {093026} (\bibinfo {year}
  {2016})}\BibitemShut {NoStop}%
\bibitem [{\citenamefont {Johansen}(2004)}]{johansen2004weak}%
  \BibitemOpen
  \bibfield  {author} {\bibinfo {author} {\bibfnamefont {L.~M.}\ \bibnamefont
  {Johansen}},\ }\href@noop {} {\bibfield  {journal} {\bibinfo  {journal}
  {Physical review letters}\ }\textbf {\bibinfo {volume} {93}},\ \bibinfo
  {pages} {120402} (\bibinfo {year} {2004})}\BibitemShut {NoStop}%
\bibitem [{\citenamefont {Kofman}\ \emph {et~al.}(2012)\citenamefont {Kofman},
  \citenamefont {Ashhab},\ and\ \citenamefont
  {Nori}}]{kofman2012nonperturbative}%
  \BibitemOpen
  \bibfield  {author} {\bibinfo {author} {\bibfnamefont {A.~G.}\ \bibnamefont
  {Kofman}}, \bibinfo {author} {\bibfnamefont {S.}~\bibnamefont {Ashhab}}, \
  and\ \bibinfo {author} {\bibfnamefont {F.}~\bibnamefont {Nori}},\ }\href@noop
  {} {\bibfield  {journal} {\bibinfo  {journal} {Physics Reports}\ }\textbf
  {\bibinfo {volume} {520}},\ \bibinfo {pages} {43} (\bibinfo {year}
  {2012})}\BibitemShut {NoStop}%
\bibitem [{\citenamefont {Lundeen}\ \emph {et~al.}(2011)\citenamefont
  {Lundeen}, \citenamefont {Sutherland}, \citenamefont {Patel}, \citenamefont
  {Stewart},\ and\ \citenamefont {Bamber}}]{lundeen2011direct}%
  \BibitemOpen
  \bibfield  {author} {\bibinfo {author} {\bibfnamefont {J.~S.}\ \bibnamefont
  {Lundeen}}, \bibinfo {author} {\bibfnamefont {B.}~\bibnamefont {Sutherland}},
  \bibinfo {author} {\bibfnamefont {A.}~\bibnamefont {Patel}}, \bibinfo
  {author} {\bibfnamefont {C.}~\bibnamefont {Stewart}}, \ and\ \bibinfo
  {author} {\bibfnamefont {C.}~\bibnamefont {Bamber}},\ }\href@noop {}
  {\bibfield  {journal} {\bibinfo  {journal} {Nature}\ }\textbf {\bibinfo
  {volume} {474}},\ \bibinfo {pages} {188} (\bibinfo {year}
  {2011})}\BibitemShut {NoStop}%
\bibitem [{\citenamefont {Lundeen}\ and\ \citenamefont
  {Bamber}(2012)}]{lundeen2012procedure}%
  \BibitemOpen
  \bibfield  {author} {\bibinfo {author} {\bibfnamefont {J.~S.}\ \bibnamefont
  {Lundeen}}\ and\ \bibinfo {author} {\bibfnamefont {C.}~\bibnamefont
  {Bamber}},\ }\href@noop {} {\bibfield  {journal} {\bibinfo  {journal}
  {Physical review letters}\ }\textbf {\bibinfo {volume} {108}},\ \bibinfo
  {pages} {070402} (\bibinfo {year} {2012})}\BibitemShut {NoStop}%
\bibitem [{\citenamefont {Thekkadath}\ \emph {et~al.}(2016)\citenamefont
  {Thekkadath}, \citenamefont {Giner}, \citenamefont {Chalich}, \citenamefont
  {Horton}, \citenamefont {Banker},\ and\ \citenamefont
  {Lundeen}}]{PhysRevLett.117.120401}%
  \BibitemOpen
  \bibfield  {author} {\bibinfo {author} {\bibfnamefont {G.~S.}\ \bibnamefont
  {Thekkadath}}, \bibinfo {author} {\bibfnamefont {L.}~\bibnamefont {Giner}},
  \bibinfo {author} {\bibfnamefont {Y.}~\bibnamefont {Chalich}}, \bibinfo
  {author} {\bibfnamefont {M.~J.}\ \bibnamefont {Horton}}, \bibinfo {author}
  {\bibfnamefont {J.}~\bibnamefont {Banker}}, \ and\ \bibinfo {author}
  {\bibfnamefont {J.~S.}\ \bibnamefont {Lundeen}},\ }\href {\doibase
  10.1103/PhysRevLett.117.120401} {\bibfield  {journal} {\bibinfo  {journal}
  {Phys. Rev. Lett.}\ }\textbf {\bibinfo {volume} {117}},\ \bibinfo {pages}
  {120401} (\bibinfo {year} {2016})}\BibitemShut {NoStop}%
\bibitem [{\citenamefont {Wu}(2013)}]{wu2013state}%
  \BibitemOpen
  \bibfield  {author} {\bibinfo {author} {\bibfnamefont {S.}~\bibnamefont
  {Wu}},\ }\href@noop {} {\bibfield  {journal} {\bibinfo  {journal} {Scientific
  reports}\ }\textbf {\bibinfo {volume} {3}},\ \bibinfo {pages} {1193}
  (\bibinfo {year} {2013})}\BibitemShut {NoStop}%
\bibitem [{\citenamefont {Vogel}\ and\ \citenamefont
  {Risken}(1989)}]{vogel1989determination}%
  \BibitemOpen
  \bibfield  {author} {\bibinfo {author} {\bibfnamefont {K.}~\bibnamefont
  {Vogel}}\ and\ \bibinfo {author} {\bibfnamefont {H.}~\bibnamefont {Risken}},\
  }\href@noop {} {\bibfield  {journal} {\bibinfo  {journal} {Physical Review
  A}\ }\textbf {\bibinfo {volume} {40}},\ \bibinfo {pages} {2847} (\bibinfo
  {year} {1989})}\BibitemShut {NoStop}%
\bibitem [{\citenamefont {Cramer}\ \emph {et~al.}(2010)\citenamefont {Cramer},
  \citenamefont {Plenio}, \citenamefont {Flammia}, \citenamefont {Somma},
  \citenamefont {Gross}, \citenamefont {Bartlett}, \citenamefont
  {Landon-Cardinal}, \citenamefont {Poulin},\ and\ \citenamefont
  {Liu}}]{cramer2010efficient}%
  \BibitemOpen
  \bibfield  {author} {\bibinfo {author} {\bibfnamefont {M.}~\bibnamefont
  {Cramer}}, \bibinfo {author} {\bibfnamefont {M.~B.}\ \bibnamefont {Plenio}},
  \bibinfo {author} {\bibfnamefont {S.~T.}\ \bibnamefont {Flammia}}, \bibinfo
  {author} {\bibfnamefont {R.}~\bibnamefont {Somma}}, \bibinfo {author}
  {\bibfnamefont {D.}~\bibnamefont {Gross}}, \bibinfo {author} {\bibfnamefont
  {S.~D.}\ \bibnamefont {Bartlett}}, \bibinfo {author} {\bibfnamefont
  {O.}~\bibnamefont {Landon-Cardinal}}, \bibinfo {author} {\bibfnamefont
  {D.}~\bibnamefont {Poulin}}, \ and\ \bibinfo {author} {\bibfnamefont {Y.-K.}\
  \bibnamefont {Liu}},\ }\href@noop {} {\bibfield  {journal} {\bibinfo
  {journal} {Nature communications}\ }\textbf {\bibinfo {volume} {1}},\
  \bibinfo {pages} {149} (\bibinfo {year} {2010})}\BibitemShut {NoStop}%
\bibitem [{\citenamefont {Von~Neumann}(1955)}]{von1955mathematical}%
  \BibitemOpen
  \bibfield  {author} {\bibinfo {author} {\bibfnamefont {J.}~\bibnamefont
  {Von~Neumann}},\ }\href@noop {} {\emph {\bibinfo {title} {Mathematical
  foundations of quantum mechanics}}},\ \bibinfo {number} {2}\ (\bibinfo
  {publisher} {Princeton university press},\ \bibinfo {year}
  {1955})\BibitemShut {NoStop}%
\bibitem [{\citenamefont {Ku}(1966)}]{ku1966notes}%
  \BibitemOpen
  \bibfield  {author} {\bibinfo {author} {\bibfnamefont {H.~H.}\ \bibnamefont
  {Ku}},\ }\href@noop {} {\bibfield  {journal} {\bibinfo  {journal} {Journal of
  Research of the National Bureau of Standards}\ }\textbf {\bibinfo {volume}
  {70}} (\bibinfo {year} {1966})}\BibitemShut {NoStop}%
\bibitem [{\citenamefont {{Gupta}}\ \emph {et~al.}(2016)\citenamefont
  {{Gupta}}, \citenamefont {{You}}, \citenamefont {{Dowling}},\ and\
  \citenamefont {{Lee}}}]{2016APS..DMP.H4009G}%
  \BibitemOpen
  \bibfield  {author} {\bibinfo {author} {\bibfnamefont {M.~K.}\ \bibnamefont
  {{Gupta}}}, \bibinfo {author} {\bibfnamefont {C.}~\bibnamefont {{You}}},
  \bibinfo {author} {\bibfnamefont {J.~P.}\ \bibnamefont {{Dowling}}}, \ and\
  \bibinfo {author} {\bibfnamefont {H.}~\bibnamefont {{Lee}}},\ }in\ \href@noop
  {} {\emph {\bibinfo {booktitle} {APS Division of Atomic, Molecular and
  Optical Physics Meeting Abstracts}}}\ (\bibinfo {year} {2016})\ p.\ \bibinfo
  {pages} {H4.009}\BibitemShut {NoStop}%
\bibitem [{\citenamefont {Gupta}(2016)}]{MKGThesis16}%
  \BibitemOpen
  \bibfield  {author} {\bibinfo {author} {\bibfnamefont {M.~K.}\ \bibnamefont
  {Gupta}},\ }\emph {\bibinfo {title} {Minimizing Decoherence in Optical Fiber
  for Long Distance Quantum Communication}},\ \href
  {https://digitalcommons.lsu.edu/gradschool_dissertations/2314} {Ph.D.
  thesis},\ \bibinfo  {school} {Louisiana State University} (\bibinfo {year}
  {2016})\BibitemShut {NoStop}%
\bibitem [{\citenamefont {Modi}\ \emph {et~al.}(2012)\citenamefont {Modi},
  \citenamefont {Brodutch}, \citenamefont {Cable}, \citenamefont {Paterek},\
  and\ \citenamefont {Vedral}}]{modi2012classical}%
  \BibitemOpen
  \bibfield  {author} {\bibinfo {author} {\bibfnamefont {K.}~\bibnamefont
  {Modi}}, \bibinfo {author} {\bibfnamefont {A.}~\bibnamefont {Brodutch}},
  \bibinfo {author} {\bibfnamefont {H.}~\bibnamefont {Cable}}, \bibinfo
  {author} {\bibfnamefont {T.}~\bibnamefont {Paterek}}, \ and\ \bibinfo
  {author} {\bibfnamefont {V.}~\bibnamefont {Vedral}},\ }\href@noop {}
  {\bibfield  {journal} {\bibinfo  {journal} {Reviews of Modern Physics}\
  }\textbf {\bibinfo {volume} {84}},\ \bibinfo {pages} {1655} (\bibinfo {year}
  {2012})}\BibitemShut {NoStop}%
\bibitem [{\citenamefont {Pande}\ and\ \citenamefont
  {Shaji}(2017)}]{pande2017minimum}%
  \BibitemOpen
  \bibfield  {author} {\bibinfo {author} {\bibfnamefont {V.~R.}\ \bibnamefont
  {Pande}}\ and\ \bibinfo {author} {\bibfnamefont {A.}~\bibnamefont {Shaji}},\
  }\href@noop {} {\bibfield  {journal} {\bibinfo  {journal} {Physics Letters
  A}\ }\textbf {\bibinfo {volume} {381}},\ \bibinfo {pages} {2045} (\bibinfo
  {year} {2017})}\BibitemShut {NoStop}%
\bibitem [{\citenamefont {Maccone}\ and\ \citenamefont
  {Rusconi}(2014)}]{maccone2014state}%
  \BibitemOpen
  \bibfield  {author} {\bibinfo {author} {\bibfnamefont {L.}~\bibnamefont
  {Maccone}}\ and\ \bibinfo {author} {\bibfnamefont {C.~C.}\ \bibnamefont
  {Rusconi}},\ }\href@noop {} {\bibfield  {journal} {\bibinfo  {journal}
  {Physical Review A}\ }\textbf {\bibinfo {volume} {89}},\ \bibinfo {pages}
  {022122} (\bibinfo {year} {2014})}\BibitemShut {NoStop}%
\bibitem [{\citenamefont {Nielsen}\ \emph {et~al.}(1998)\citenamefont
  {Nielsen}, \citenamefont {Knill},\ and\ \citenamefont
  {Laflamme}}]{nielsen1998complete}%
  \BibitemOpen
  \bibfield  {author} {\bibinfo {author} {\bibfnamefont {M.}~\bibnamefont
  {Nielsen}}, \bibinfo {author} {\bibfnamefont {E.}~\bibnamefont {Knill}}, \
  and\ \bibinfo {author} {\bibfnamefont {R.}~\bibnamefont {Laflamme}},\
  }\href@noop {} {\bibfield  {journal} {\bibinfo  {journal} {Nature}\ }\textbf
  {\bibinfo {volume} {396}},\ \bibinfo {pages} {52} (\bibinfo {year}
  {1998})}\BibitemShut {NoStop}%
\bibitem [{\citenamefont {Xu}\ \emph {et~al.}(2010)\citenamefont {Xu},
  \citenamefont {Xu}, \citenamefont {Li}, \citenamefont {Zhang}, \citenamefont
  {Zou},\ and\ \citenamefont {Guo}}]{xu2010experimental}%
  \BibitemOpen
  \bibfield  {author} {\bibinfo {author} {\bibfnamefont {J.-S.}\ \bibnamefont
  {Xu}}, \bibinfo {author} {\bibfnamefont {X.-Y.}\ \bibnamefont {Xu}}, \bibinfo
  {author} {\bibfnamefont {C.-F.}\ \bibnamefont {Li}}, \bibinfo {author}
  {\bibfnamefont {C.-J.}\ \bibnamefont {Zhang}}, \bibinfo {author}
  {\bibfnamefont {X.-B.}\ \bibnamefont {Zou}}, \ and\ \bibinfo {author}
  {\bibfnamefont {G.-C.}\ \bibnamefont {Guo}},\ }\href@noop {} {\bibfield
  {journal} {\bibinfo  {journal} {Nature communications}\ }\textbf {\bibinfo
  {volume} {1}},\ \bibinfo {pages} {7} (\bibinfo {year} {2010})}\BibitemShut
  {NoStop}%
\bibitem [{\citenamefont {Bae}\ \emph {et~al.}(2005)\citenamefont {Bae},
  \citenamefont {Jin}, \citenamefont {Kim}, \citenamefont {Yoon},\ and\
  \citenamefont {Kwon}}]{bae2005three}%
  \BibitemOpen
  \bibfield  {author} {\bibinfo {author} {\bibfnamefont {J.}~\bibnamefont
  {Bae}}, \bibinfo {author} {\bibfnamefont {J.}~\bibnamefont {Jin}}, \bibinfo
  {author} {\bibfnamefont {J.}~\bibnamefont {Kim}}, \bibinfo {author}
  {\bibfnamefont {C.}~\bibnamefont {Yoon}}, \ and\ \bibinfo {author}
  {\bibfnamefont {Y.}~\bibnamefont {Kwon}},\ }\href@noop {} {\bibfield
  {journal} {\bibinfo  {journal} {Chaos, Solitons \& Fractals}\ }\textbf
  {\bibinfo {volume} {24}},\ \bibinfo {pages} {1047} (\bibinfo {year}
  {2005})}\BibitemShut {NoStop}%
\bibitem [{\citenamefont {Pirandola}\ \emph {et~al.}(2005)\citenamefont
  {Pirandola}, \citenamefont {Mancini},\ and\ \citenamefont
  {Vitali}}]{pirandola2005conditioning}%
  \BibitemOpen
  \bibfield  {author} {\bibinfo {author} {\bibfnamefont {S.}~\bibnamefont
  {Pirandola}}, \bibinfo {author} {\bibfnamefont {S.}~\bibnamefont {Mancini}},
  \ and\ \bibinfo {author} {\bibfnamefont {D.}~\bibnamefont {Vitali}},\
  }\href@noop {} {\bibfield  {journal} {\bibinfo  {journal} {Physical Review
  A}\ }\textbf {\bibinfo {volume} {71}},\ \bibinfo {pages} {042326} (\bibinfo
  {year} {2005})}\BibitemShut {NoStop}%
\bibitem [{\citenamefont {Hu}\ \emph {et~al.}(2016)\citenamefont {Hu},
  \citenamefont {Zhou}, \citenamefont {Hu}, \citenamefont {Li}, \citenamefont
  {Guo},\ and\ \citenamefont {Zhang}}]{hu2016experimental}%
  \BibitemOpen
  \bibfield  {author} {\bibinfo {author} {\bibfnamefont {M.-J.}\ \bibnamefont
  {Hu}}, \bibinfo {author} {\bibfnamefont {Z.-Y.}\ \bibnamefont {Zhou}},
  \bibinfo {author} {\bibfnamefont {X.-M.}\ \bibnamefont {Hu}}, \bibinfo
  {author} {\bibfnamefont {C.-F.}\ \bibnamefont {Li}}, \bibinfo {author}
  {\bibfnamefont {G.-C.}\ \bibnamefont {Guo}}, \ and\ \bibinfo {author}
  {\bibfnamefont {Y.-S.}\ \bibnamefont {Zhang}},\ }\href@noop {} {\bibfield
  {journal} {\bibinfo  {journal} {arXiv preprint arXiv:1609.01863}\ } (\bibinfo
  {year} {2016})}\BibitemShut {NoStop}%
\bibitem [{\citenamefont {Lo}(2000)}]{lo2000classical}%
  \BibitemOpen
  \bibfield  {author} {\bibinfo {author} {\bibfnamefont {H.-K.}\ \bibnamefont
  {Lo}},\ }\href@noop {} {\bibfield  {journal} {\bibinfo  {journal} {Physical
  Review A}\ }\textbf {\bibinfo {volume} {62}},\ \bibinfo {pages} {012313}
  (\bibinfo {year} {2000})}\BibitemShut {NoStop}%
\bibitem [{\citenamefont {Gisin}\ and\ \citenamefont
  {Thew}(2007)}]{gisin2007quantum}%
  \BibitemOpen
  \bibfield  {author} {\bibinfo {author} {\bibfnamefont {N.}~\bibnamefont
  {Gisin}}\ and\ \bibinfo {author} {\bibfnamefont {R.}~\bibnamefont {Thew}},\
  }\href@noop {} {\bibfield  {journal} {\bibinfo  {journal} {Nature photonics}\
  }\textbf {\bibinfo {volume} {1}},\ \bibinfo {pages} {165} (\bibinfo {year}
  {2007})}\BibitemShut {NoStop}%
\bibitem [{\citenamefont {Nielsen}\ and\ \citenamefont
  {Chuang}(2010)}]{nielsen2010quantum}%
  \BibitemOpen
  \bibfield  {author} {\bibinfo {author} {\bibfnamefont {M.~A.}\ \bibnamefont
  {Nielsen}}\ and\ \bibinfo {author} {\bibfnamefont {I.~L.}\ \bibnamefont
  {Chuang}},\ }\href@noop {} {\bibfield  {journal} {\bibinfo  {journal}
  {Quantum Computation and Quantum Information, by Michael A. Nielsen, Isaac L.
  Chuang, Cambridge, UK: Cambridge University Press, 2010}\ } (\bibinfo {year}
  {2010})}\BibitemShut {NoStop}%
\bibitem [{\citenamefont {Roszak}\ \emph {et~al.}(2015)\citenamefont {Roszak}
  \emph {et~al.}}]{roszak2015relation}%
  \BibitemOpen
  \bibfield  {author} {\bibinfo {author} {\bibfnamefont {K.}~\bibnamefont
  {Roszak}} \emph {et~al.},\ }\href@noop {} {\bibfield  {journal} {\bibinfo
  {journal} {EPL (Europhysics Letters)}\ }\textbf {\bibinfo {volume} {112}},\
  \bibinfo {pages} {10002} (\bibinfo {year} {2015})}\BibitemShut {NoStop}%
\bibitem [{\citenamefont {Wiseman}\ \emph {et~al.}(2007)\citenamefont
  {Wiseman}, \citenamefont {Jones},\ and\ \citenamefont
  {Doherty}}]{wiseman2007steering}%
  \BibitemOpen
  \bibfield  {author} {\bibinfo {author} {\bibfnamefont {H.~M.}\ \bibnamefont
  {Wiseman}}, \bibinfo {author} {\bibfnamefont {S.~J.}\ \bibnamefont {Jones}},
  \ and\ \bibinfo {author} {\bibfnamefont {A.~C.}\ \bibnamefont {Doherty}},\
  }\href@noop {} {\bibfield  {journal} {\bibinfo  {journal} {Physical review
  letters}\ }\textbf {\bibinfo {volume} {98}},\ \bibinfo {pages} {140402}
  (\bibinfo {year} {2007})}\BibitemShut {NoStop}%
\bibitem [{\citenamefont {Buscemi}(2012)}]{buscemi2012all}%
  \BibitemOpen
  \bibfield  {author} {\bibinfo {author} {\bibfnamefont {F.}~\bibnamefont
  {Buscemi}},\ }\href@noop {} {\bibfield  {journal} {\bibinfo  {journal}
  {Physical review letters}\ }\textbf {\bibinfo {volume} {108}},\ \bibinfo
  {pages} {200401} (\bibinfo {year} {2012})}\BibitemShut {NoStop}%
\bibitem [{\citenamefont {Guo}\ and\ \citenamefont
  {Wu}(2014)}]{guo2014quantum}%
  \BibitemOpen
  \bibfield  {author} {\bibinfo {author} {\bibfnamefont {Y.}~\bibnamefont
  {Guo}}\ and\ \bibinfo {author} {\bibfnamefont {S.}~\bibnamefont {Wu}},\
  }\href@noop {} {\bibfield  {journal} {\bibinfo  {journal} {Scientific
  reports}\ }\textbf {\bibinfo {volume} {4}} (\bibinfo {year}
  {2014})}\BibitemShut {NoStop}%
\bibitem [{\citenamefont {Masanes}\ \emph {et~al.}(2008)\citenamefont
  {Masanes}, \citenamefont {Liang},\ and\ \citenamefont
  {Doherty}}]{masanes2008all}%
  \BibitemOpen
  \bibfield  {author} {\bibinfo {author} {\bibfnamefont {L.}~\bibnamefont
  {Masanes}}, \bibinfo {author} {\bibfnamefont {Y.-C.}\ \bibnamefont {Liang}},
  \ and\ \bibinfo {author} {\bibfnamefont {A.~C.}\ \bibnamefont {Doherty}},\
  }\href@noop {} {\bibfield  {journal} {\bibinfo  {journal} {Physical review
  letters}\ }\textbf {\bibinfo {volume} {100}},\ \bibinfo {pages} {090403}
  (\bibinfo {year} {2008})}\BibitemShut {NoStop}%
\bibitem [{\citenamefont {Peres}(1996)}]{peres1996separability}%
  \BibitemOpen
  \bibfield  {author} {\bibinfo {author} {\bibfnamefont {A.}~\bibnamefont
  {Peres}},\ }\href@noop {} {\bibfield  {journal} {\bibinfo  {journal}
  {Physical Review Letters}\ }\textbf {\bibinfo {volume} {77}},\ \bibinfo
  {pages} {1413} (\bibinfo {year} {1996})}\BibitemShut {NoStop}%
\bibitem [{\citenamefont {Maccone}\ \emph {et~al.}(2015)\citenamefont
  {Maccone}, \citenamefont {Bruss},\ and\ \citenamefont
  {Macchiavello}}]{maccone2015complementarity}%
  \BibitemOpen
  \bibfield  {author} {\bibinfo {author} {\bibfnamefont {L.}~\bibnamefont
  {Maccone}}, \bibinfo {author} {\bibfnamefont {D.}~\bibnamefont {Bruss}}, \
  and\ \bibinfo {author} {\bibfnamefont {C.}~\bibnamefont {Macchiavello}},\
  }\href@noop {} {\bibfield  {journal} {\bibinfo  {journal} {Physical review
  letters}\ }\textbf {\bibinfo {volume} {114}},\ \bibinfo {pages} {130401}
  (\bibinfo {year} {2015})}\BibitemShut {NoStop}%
\bibitem [{\citenamefont {Ac\'{\i}n}\ \emph {et~al.}(2010)\citenamefont
  {Ac\'{\i}n}, \citenamefont {Augusiak}, \citenamefont {Cavalcanti},
  \citenamefont {Hadley}, \citenamefont {Korbicz}, \citenamefont {Lewenstein},
  \citenamefont {Masanes},\ and\ \citenamefont
  {Piani}}]{PhysRevLett.104.140404}%
  \BibitemOpen
  \bibfield  {author} {\bibinfo {author} {\bibfnamefont {A.}~\bibnamefont
  {Ac\'{\i}n}}, \bibinfo {author} {\bibfnamefont {R.}~\bibnamefont {Augusiak}},
  \bibinfo {author} {\bibfnamefont {D.}~\bibnamefont {Cavalcanti}}, \bibinfo
  {author} {\bibfnamefont {C.}~\bibnamefont {Hadley}}, \bibinfo {author}
  {\bibfnamefont {J.~K.}\ \bibnamefont {Korbicz}}, \bibinfo {author}
  {\bibfnamefont {M.}~\bibnamefont {Lewenstein}}, \bibinfo {author}
  {\bibfnamefont {L.}~\bibnamefont {Masanes}}, \ and\ \bibinfo {author}
  {\bibfnamefont {M.}~\bibnamefont {Piani}},\ }\href {\doibase
  10.1103/PhysRevLett.104.140404} {\bibfield  {journal} {\bibinfo  {journal}
  {Phys. Rev. Lett.}\ }\textbf {\bibinfo {volume} {104}},\ \bibinfo {pages}
  {140404} (\bibinfo {year} {2010})}\BibitemShut {NoStop}%
\bibitem [{\citenamefont {Bartlett}\ \emph {et~al.}(2006)\citenamefont
  {Bartlett}, \citenamefont {Doherty}, \citenamefont {Spekkens},\ and\
  \citenamefont {Wiseman}}]{PhysRevA.73.022311}%
  \BibitemOpen
  \bibfield  {author} {\bibinfo {author} {\bibfnamefont {S.~D.}\ \bibnamefont
  {Bartlett}}, \bibinfo {author} {\bibfnamefont {A.~C.}\ \bibnamefont
  {Doherty}}, \bibinfo {author} {\bibfnamefont {R.~W.}\ \bibnamefont
  {Spekkens}}, \ and\ \bibinfo {author} {\bibfnamefont {H.~M.}\ \bibnamefont
  {Wiseman}},\ }\href {\doibase 10.1103/PhysRevA.73.022311} {\bibfield
  {journal} {\bibinfo  {journal} {Phys. Rev. A}\ }\textbf {\bibinfo {volume}
  {73}},\ \bibinfo {pages} {022311} (\bibinfo {year} {2006})}\BibitemShut
  {NoStop}%
\bibitem [{\citenamefont {Ferraro}\ \emph {et~al.}(2010)\citenamefont
  {Ferraro}, \citenamefont {Aolita}, \citenamefont {Cavalcanti}, \citenamefont
  {Cucchietti},\ and\ \citenamefont {Ac\'{\i}n}}]{PhysRevA.81.052318}%
  \BibitemOpen
  \bibfield  {author} {\bibinfo {author} {\bibfnamefont {A.}~\bibnamefont
  {Ferraro}}, \bibinfo {author} {\bibfnamefont {L.}~\bibnamefont {Aolita}},
  \bibinfo {author} {\bibfnamefont {D.}~\bibnamefont {Cavalcanti}}, \bibinfo
  {author} {\bibfnamefont {F.~M.}\ \bibnamefont {Cucchietti}}, \ and\ \bibinfo
  {author} {\bibfnamefont {A.}~\bibnamefont {Ac\'{\i}n}},\ }\href {\doibase
  10.1103/PhysRevA.81.052318} {\bibfield  {journal} {\bibinfo  {journal} {Phys.
  Rev. A}\ }\textbf {\bibinfo {volume} {81}},\ \bibinfo {pages} {052318}
  (\bibinfo {year} {2010})}\BibitemShut {NoStop}%
\bibitem [{\citenamefont {Bell}(1964)}]{bell1964einstein}%
  \BibitemOpen
  \bibfield  {author} {\bibinfo {author} {\bibfnamefont {J.~S.}\ \bibnamefont
  {Bell}},\ }\href@noop {} {\enquote {\bibinfo {title} {On the einstein
  podolsky rosen paradox},}\ } (\bibinfo {year} {1964})\BibitemShut {NoStop}%
\bibitem [{\citenamefont {Werner}(1989)}]{werner1989quantum}%
  \BibitemOpen
  \bibfield  {author} {\bibinfo {author} {\bibfnamefont {R.~F.}\ \bibnamefont
  {Werner}},\ }\href@noop {} {\bibfield  {journal} {\bibinfo  {journal}
  {Physical Review A}\ }\textbf {\bibinfo {volume} {40}},\ \bibinfo {pages}
  {4277} (\bibinfo {year} {1989})}\BibitemShut {NoStop}%
\bibitem [{\citenamefont {Silva}\ \emph {et~al.}(2015)\citenamefont {Silva},
  \citenamefont {Gisin}, \citenamefont {Guryanova},\ and\ \citenamefont
  {Popescu}}]{silva2015multiple}%
  \BibitemOpen
  \bibfield  {author} {\bibinfo {author} {\bibfnamefont {R.}~\bibnamefont
  {Silva}}, \bibinfo {author} {\bibfnamefont {N.}~\bibnamefont {Gisin}},
  \bibinfo {author} {\bibfnamefont {Y.}~\bibnamefont {Guryanova}}, \ and\
  \bibinfo {author} {\bibfnamefont {S.}~\bibnamefont {Popescu}},\ }\href@noop
  {} {\bibfield  {journal} {\bibinfo  {journal} {Physical review letters}\
  }\textbf {\bibinfo {volume} {114}},\ \bibinfo {pages} {250401} (\bibinfo
  {year} {2015})}\BibitemShut {NoStop}%
\bibitem [{\citenamefont {Gisin}(1991)}]{gisin1991bell}%
  \BibitemOpen
  \bibfield  {author} {\bibinfo {author} {\bibfnamefont {N.}~\bibnamefont
  {Gisin}},\ }\href@noop {} {\bibfield  {journal} {\bibinfo  {journal} {Physics
  Letters A}\ }\textbf {\bibinfo {volume} {154}},\ \bibinfo {pages} {201}
  (\bibinfo {year} {1991})}\BibitemShut {NoStop}%
\bibitem [{\citenamefont {Lanyon}\ \emph {et~al.}(2008)\citenamefont {Lanyon},
  \citenamefont {Barbieri}, \citenamefont {Almeida},\ and\ \citenamefont
  {White}}]{lanyon2008experimental}%
  \BibitemOpen
  \bibfield  {author} {\bibinfo {author} {\bibfnamefont {B.~P.}\ \bibnamefont
  {Lanyon}}, \bibinfo {author} {\bibfnamefont {M.}~\bibnamefont {Barbieri}},
  \bibinfo {author} {\bibfnamefont {M.~P.}\ \bibnamefont {Almeida}}, \ and\
  \bibinfo {author} {\bibfnamefont {A.~G.}\ \bibnamefont {White}},\ }\href@noop
  {} {\bibfield  {journal} {\bibinfo  {journal} {Physical Review Letters}\
  }\textbf {\bibinfo {volume} {101}} (\bibinfo {year} {2008})}\BibitemShut
  {NoStop}%
\bibitem [{\citenamefont {Wang}\ \emph {et~al.}(2013)\citenamefont {Wang},
  \citenamefont {Huang}, \citenamefont {Dowling},\ and\ \citenamefont
  {Zhu}}]{wang2012quantum}%
  \BibitemOpen
  \bibfield  {author} {\bibinfo {author} {\bibfnamefont {L.}~\bibnamefont
  {Wang}}, \bibinfo {author} {\bibfnamefont {J.-H.}\ \bibnamefont {Huang}},
  \bibinfo {author} {\bibfnamefont {J.~P.}\ \bibnamefont {Dowling}}, \ and\
  \bibinfo {author} {\bibfnamefont {S.-Y.}\ \bibnamefont {Zhu}},\ }\href@noop
  {} {\bibfield  {journal} {\bibinfo  {journal} {Quantum Information
  Processing}\ }\textbf {\bibinfo {volume} {12}},\ \bibinfo {pages} {899}
  (\bibinfo {year} {2013})}\BibitemShut {NoStop}%
\bibitem [{\citenamefont {Grosshans}\ and\ \citenamefont
  {Grangier}(2001)}]{PhysRevA.64.010301}%
  \BibitemOpen
  \bibfield  {author} {\bibinfo {author} {\bibfnamefont {F.}~\bibnamefont
  {Grosshans}}\ and\ \bibinfo {author} {\bibfnamefont {P.}~\bibnamefont
  {Grangier}},\ }\href {\doibase 10.1103/PhysRevA.64.010301} {\bibfield
  {journal} {\bibinfo  {journal} {Phys. Rev. A}\ }\textbf {\bibinfo {volume}
  {64}},\ \bibinfo {pages} {010301} (\bibinfo {year} {2001})}\BibitemShut
  {NoStop}%
\bibitem [{\citenamefont {Coto}\ \emph {et~al.}(2017)\citenamefont {Coto},
  \citenamefont {Montenegro}, \citenamefont {Eremeev}, \citenamefont
  {Mundarain},\ and\ \citenamefont {Orszag}}]{coto2017power}%
  \BibitemOpen
  \bibfield  {author} {\bibinfo {author} {\bibfnamefont {R.}~\bibnamefont
  {Coto}}, \bibinfo {author} {\bibfnamefont {V.}~\bibnamefont {Montenegro}},
  \bibinfo {author} {\bibfnamefont {V.}~\bibnamefont {Eremeev}}, \bibinfo
  {author} {\bibfnamefont {D.}~\bibnamefont {Mundarain}}, \ and\ \bibinfo
  {author} {\bibfnamefont {M.}~\bibnamefont {Orszag}},\ }\href@noop {}
  {\bibfield  {journal} {\bibinfo  {journal} {Scientific reports}\ }\textbf
  {\bibinfo {volume} {7}},\ \bibinfo {pages} {6351} (\bibinfo {year}
  {2017})}\BibitemShut {NoStop}%
\bibitem [{\citenamefont {Dressel}\ \emph {et~al.}(2014)\citenamefont
  {Dressel}, \citenamefont {Malik}, \citenamefont {Miatto}, \citenamefont
  {Jordan},\ and\ \citenamefont {Boyd}}]{dressel2014colloquium}%
  \BibitemOpen
  \bibfield  {author} {\bibinfo {author} {\bibfnamefont {J.}~\bibnamefont
  {Dressel}}, \bibinfo {author} {\bibfnamefont {M.}~\bibnamefont {Malik}},
  \bibinfo {author} {\bibfnamefont {F.~M.}\ \bibnamefont {Miatto}}, \bibinfo
  {author} {\bibfnamefont {A.~N.}\ \bibnamefont {Jordan}}, \ and\ \bibinfo
  {author} {\bibfnamefont {R.~W.}\ \bibnamefont {Boyd}},\ }\href@noop {}
  {\bibfield  {journal} {\bibinfo  {journal} {Reviews of Modern Physics}\
  }\textbf {\bibinfo {volume} {86}},\ \bibinfo {pages} {307} (\bibinfo {year}
  {2014})}\BibitemShut {NoStop}%
\bibitem [{\citenamefont {Mirhosseini}\ \emph {et~al.}(2014)\citenamefont
  {Mirhosseini}, \citenamefont {Magaña-Loaiza}, \citenamefont
  {Hashemi~Rafsanjani},\ and\ \citenamefont
  {Boyd}}]{mirhosseini2014compressive}%
  \BibitemOpen
  \bibfield  {author} {\bibinfo {author} {\bibfnamefont {M.}~\bibnamefont
  {Mirhosseini}}, \bibinfo {author} {\bibfnamefont {O.~S.}\ \bibnamefont
  {Magaña-Loaiza}}, \bibinfo {author} {\bibfnamefont {S.~M.}\ \bibnamefont
  {Hashemi~Rafsanjani}}, \ and\ \bibinfo {author} {\bibfnamefont {R.~W.}\
  \bibnamefont {Boyd}},\ }\href@noop {} {\bibfield  {journal} {\bibinfo
  {journal} {Physical Review Letters}\ }\textbf {\bibinfo {volume} {113}}
  (\bibinfo {year} {2014})}\BibitemShut {NoStop}%
\bibitem [{\citenamefont {Zhou}\ \emph {et~al.}(2021)\citenamefont {Zhou},
  \citenamefont {Zhao}, \citenamefont {Hay}, \citenamefont {McGonagle},
  \citenamefont {Boyd},\ and\ \citenamefont {Shi}}]{PhysRevLett.127.040402}%
  \BibitemOpen
  \bibfield  {author} {\bibinfo {author} {\bibfnamefont {Y.}~\bibnamefont
  {Zhou}}, \bibinfo {author} {\bibfnamefont {J.}~\bibnamefont {Zhao}}, \bibinfo
  {author} {\bibfnamefont {D.}~\bibnamefont {Hay}}, \bibinfo {author}
  {\bibfnamefont {K.}~\bibnamefont {McGonagle}}, \bibinfo {author}
  {\bibfnamefont {R.~W.}\ \bibnamefont {Boyd}}, \ and\ \bibinfo {author}
  {\bibfnamefont {Z.}~\bibnamefont {Shi}},\ }\href {\doibase
  10.1103/PhysRevLett.127.040402} {\bibfield  {journal} {\bibinfo  {journal}
  {Phys. Rev. Lett.}\ }\textbf {\bibinfo {volume} {127}},\ \bibinfo {pages}
  {040402} (\bibinfo {year} {2021})}\BibitemShut {NoStop}%
\bibitem [{\citenamefont {Resch}\ and\ \citenamefont
  {Steinberg}(2004)}]{PhysRevLett.92.130402}%
  \BibitemOpen
  \bibfield  {author} {\bibinfo {author} {\bibfnamefont {K.~J.}\ \bibnamefont
  {Resch}}\ and\ \bibinfo {author} {\bibfnamefont {A.~M.}\ \bibnamefont
  {Steinberg}},\ }\href {\doibase 10.1103/PhysRevLett.92.130402} {\bibfield
  {journal} {\bibinfo  {journal} {Phys. Rev. Lett.}\ }\textbf {\bibinfo
  {volume} {92}},\ \bibinfo {pages} {130402} (\bibinfo {year}
  {2004})}\BibitemShut {NoStop}%
\bibitem [{\citenamefont {Puentes}\ \emph {et~al.}(2012)\citenamefont
  {Puentes}, \citenamefont {Hermosa},\ and\ \citenamefont
  {Torres}}]{PhysRevLett.109.040401}%
  \BibitemOpen
  \bibfield  {author} {\bibinfo {author} {\bibfnamefont {G.}~\bibnamefont
  {Puentes}}, \bibinfo {author} {\bibfnamefont {N.}~\bibnamefont {Hermosa}}, \
  and\ \bibinfo {author} {\bibfnamefont {J.~P.}\ \bibnamefont {Torres}},\
  }\href {\doibase 10.1103/PhysRevLett.109.040401} {\bibfield  {journal}
  {\bibinfo  {journal} {Phys. Rev. Lett.}\ }\textbf {\bibinfo {volume} {109}},\
  \bibinfo {pages} {040401} (\bibinfo {year} {2012})}\BibitemShut {NoStop}%
\bibitem [{\citenamefont {Pati}\ and\ \citenamefont
  {Singh}(2013)}]{pati2013weak}%
  \BibitemOpen
  \bibfield  {author} {\bibinfo {author} {\bibfnamefont {A.~K.}\ \bibnamefont
  {Pati}}\ and\ \bibinfo {author} {\bibfnamefont {U.}~\bibnamefont {Singh}},\
  }\href@noop {} {\bibfield  {journal} {\bibinfo  {journal} {arXiv preprint
  arXiv:1310.6002}\ } (\bibinfo {year} {2013})}\BibitemShut {NoStop}%
\bibitem [{\citenamefont {Bertlmann}\ and\ \citenamefont
  {Krammer}(2008)}]{bertlmann2008bloch}%
  \BibitemOpen
  \bibfield  {author} {\bibinfo {author} {\bibfnamefont {R.~A.}\ \bibnamefont
  {Bertlmann}}\ and\ \bibinfo {author} {\bibfnamefont {P.}~\bibnamefont
  {Krammer}},\ }\href@noop {} {\bibfield  {journal} {\bibinfo  {journal}
  {Journal of Physics A: Mathematical and Theoretical}\ }\textbf {\bibinfo
  {volume} {41}},\ \bibinfo {pages} {235303} (\bibinfo {year}
  {2008})}\BibitemShut {NoStop}%
\bibitem [{\citenamefont {Stephany}(1979)}]{stephany1979higher}%
  \BibitemOpen
  \bibfield  {author} {\bibinfo {author} {\bibfnamefont {J.}~\bibnamefont
  {Stephany}},\ }\href@noop {} {\bibfield  {journal} {\bibinfo  {journal}
  {Journal of Physics A: Mathematical and General}\ }\textbf {\bibinfo {volume}
  {12}},\ \bibinfo {pages} {1667} (\bibinfo {year} {1979})}\BibitemShut
  {NoStop}%
\end{thebibliography}%

\end{document}